\documentclass[12pt]{article}
\def\be{\begin{equation}} \def\ee{\end{equation}} \def\bea{\begin{eqnarray}}
\def\eea{\end{eqnarray}} \def\nnb{\nonumber}

\usepackage{graphicx}

\begin{document}

\hfill{October 4, 2026,\ \ \ {\tt 13CanOv4.1}} 

\begin{center}
\vskip 5mm 

\noindent
{\Large\bf  
	$S$ factor of $^{13}$C($\alpha$,$n$)$^{16}$O 
		at low energies 
	in cluster effective field theory 
}
\vskip 6mm 

\noindent
{\large 
Shung-Ichi Ando\footnote{mailto:sando@sunmoon.ac.kr}, 
\vskip 6mm
\noindent
{\it
Department of Display and Semiconductor Engineering
	and Research Center for Nano-Bio Science, 
Sunmoon University,
Asan, Chungnam 31460,
Republic of Korea

Department of Physics Education,
Daegu University, Gyeongsan 38453,
Republic of Korea
}
	}
\end{center}

\vskip 6mm

The $^{13}$C($\alpha$,$n$)$^{16}$O reaction at low energies 
is studied by constructing an effective field theory.
We choose a separation scale at $E=1$~MeV, 
where $E$ is the initial $\alpha$-$^{13}$C energy in center-of-mass frame,
just below the sharp resonant $5/2^+$ state of $^{17}$O, 
and include two open channels, $\alpha$-$^{13}$C and $n$-$^{16}$O,
and resonant $1/2^+$, $5/2^-$, $3/2^+$ states of $^{17}$O in the study.  
Parameters of the theory are fitted to experimental data,
$S$ factor of $^{13}$C($\alpha$,$n$)$^{16}$O 
at the energies below $E=1$~MeV, 
including the data sets recently reported by the LUNA and JUNA collaborations,
and the $S$ factor of $^{13}$C($\alpha$,$n$)$^{16}$O 
is extrapolated to the Gamow peak energy $E_G= 0.19$~MeV
in the low-mass AGB stars. 
We discuss an uncertainty in the estimate of the $S$ factor 
and confirm that the main part of the uncertainty emerges from 
the model assumptions about  
the near-breakup threshold $1/2^+$ state of $^{17}$O. 

\newpage
\noindent
{\bf 1. Introduction}

The $^{13}$C($\alpha$,$n$)$^{16}$O reaction provides a main neutron 
source in the $s$ process in low-mass asymptotic giant branch (AGB) stars 
with the mass $1.5 \le M/M_\odot \le 3$~\cite{setal-npa06}. 
A $^{13}$C pocket, where $^{13}$C is produced through
$^{12}$C($p$,$\gamma$)$^{13}$N($\beta^+\nu$)$^{13}$C, is formed 
in a thin (He-rich and C-rich) He-intershell between a compact C-O core
and an expanded H-rich envelope at the epoch of the so-called third dredge up. 
The $^{13}$C($\alpha$,$n$)$^{16}$O reaction takes place during 
the interpulse phase of the AGB stars, $\sim 10^4$~yr, 
when the temperature rises up to $\sim 90\times 10^6$~K. 
It leads to a maximum neutron density of about $10^7$~cm$^{-3}$. 
By neutron capture reactions (and following $\beta$ decay reactions) 
heavier elements, $Z>30$, are synthesized through the line 
of stability. 
About half of the heavy elements in the universe
have been created in the $s$ process 
in the low-mass AGB stars~\cite{scg-araa08}. 

Recently, precise measurements of the $^{13}$C($\alpha$,$n$)$^{16}$O reaction were reported
from the LUNA collaboration~\cite{LUNA} and 
the JUNA collaboration~\cite{JUNA}, 
and the energy ranges of the measurements are
$E = 230$ to 300 keV and 240 to 1800 keV, respectively.
They are close to the Gamow peak energy,
$E_G = 190\pm 40$~keV,
for the $\alpha$-$^{13}$C system 
at $\sim 90\times 10^6$~K. 
It may be straightforward to extrapolate the cross section, equivalently 
the astrophysical $S$ factor, 
of the $^{13}$C($\alpha$,$n$)$^{16}$O 
reaction to $E_G$ with the new data, 
 but an uncertainty in the estimate of the $S$ factor
emerges from the low-energy side of the experimental data, 
due to a broad resonant $1/2^+$ state of $^{17}$O appearing
around the breakup energy of the $\alpha$-$^{13}$C channel~\cite{detal-prc20}. 
In this work, we construct 
an effective field theory (EFT) for the study of  
the $^{13}$C($\alpha$,$n$)$^{16}$O reaction, extrapolate the $S$ 
factor to $E_G$, and confirm the uncertainty from the near-threshold state. 

An EFT for the study of the reaction in question is constructed 
by introducing a separation scale between relevant degrees of freedom
at low energies and irrelevant degrees of freedom at high 
energies~\cite{w-physica79,dgh-14}.
An effective Lagrangian is constructed by requiring symmetries
and is expanded in terms of the number of derivatives on the 
fields.
The theory provides us with a perturbative expansion scheme 
in powers of ${\cal Q}/\Lambda_H$ 
where ${\cal Q}$ is a typical momentum scale 
and $\Lambda_H$ is a large momentum scale, 
and the reaction amplitudes are calculated order by order.  
The infinities from loop calculations, 
where the loops preserve the unitary condition of the amplitudes, 
are subtracted by the coefficients of the effective Lagrangian; 
the contributions from the irrelevant degrees of freedom 
at high energies are embedded in the coefficients, 
while the coefficients are practically fitted to experimental data. 
The perturbative expansion scheme in an EFT would be useful to 
estimate a systematic uncertainty due to the truncation of perturbative series. 
For the last decade, we applied the methodology of EFT to the study of 
nuclear reactions for nuclear astrophysics. 
We constructed an EFT for $\alpha$-$^{12}$C system and
studied elastic $\alpha$-$^{12}$C 
scattering~\cite{sa-epja16,sa-prc18,sa-prc23,metal-25}, 
$\beta$ delayed $\alpha$ emission from $^{16}$N~\cite{sa-epja21}, 
and $E1$ and $E2$ transitions of 
$^{12}$C($\alpha$,$\gamma$)$^{16}$O reaction~\cite{sa-prc19,sa-cpc25,sa-npa25}. 

In this work, we construct an EFT for the study of 
$^{13}$C($\alpha$,$n$)$^{16}$O reaction at low energies and extrapolate
the $S$ factor of the reaction to $E_G$. 
As we will discuss in the next section, 
we introduce three resonant $1/2^+$, $5/2^-$, $3/2^+$ states of $^{17}$O
as relevant degrees of freedom in the present study, and we employ composite 
fields for them. In the energy region close to the resonant poles, 
one needs to treat interactions non-perturbatively~\cite{hfvk-epja21}.
By employing composite fields, one can expand the terms
in the unitary limits: those terms are matched to the effective range 
parameters~\cite{k-npb97,bs-npa01,ah-prc05}. We also introduce two
open channels, $\alpha$-$^{13}$C and $n$-$^{16}$O. 
To construct projection operators of the two open channels 
to the resonant fields for the $3/2^+$ and $5/2^-$ states,
we employ the $2\times 4$ matrices for the $3/2^+$ state reported
by Fernando, Higa, and Rupak~\cite{fhr-epja12}.
For the projection operators to connect the two open channels 
to the $5/2^-$ state, we construct the $2\times 6$ matrices. 
We write down the effective Lagrangian, 
extract the vertices and propagators from the Lagrangian,
and calculate the vertices and self-energies, including 
the non-perturbative Coulomb interaction.
Then, we obtain an expression for the reaction amplitudes of 
$^{13}$C($\alpha$,$n$)$^{16}$O for the three resonant states of $^{17}$O
in EFT. 
The parameters in the theory are fitted to the available experimental
data of the $S$ factor at $0.23 \le E\le 1$~MeV, 
and the $S$ factor is extrapolated to $E_G=0.19$~MeV.
We then discuss the uncertainties of the estimate of the $S$ factor 
in this work. 

The present work is organized as follows.
In Section 2, we discuss the composition of an EFT 
for the $^{13}$C($\alpha$,$n$)$^{16}$O reaction at low energies,
and in Section 3, the effective Lagrangian for the present work
is presented. In Section 4, the expressions of $S$ factors and 
reaction amplitudes are displayed. 
In Section 5, the numerical results of this work are reported, 
and in Section 6, the results and discussions of this work are presented.  
In the appendices, we discuss the calculation of the components 
of the amplitudes in detail. 
In Appendix A, the projection operators for the $1/2^+$, $5/2^-$, $3/2^+$
states are discussed, and in Appendix B, the angular momentum projection
operators for $l=1,2,3$ between the two clusters of the open channels are
displayed. 
In Appendix C, the expressions of the components of the reaction 
amplitudes; bare propagators, vertex functions,
self-energies, and dressed propagators, are displayed and discussed.  

\vskip 2mm \noindent 
{\bf 2. Composing an EFT of $^{13}$C($\alpha$,$n$)$^{16}$O at 
low energies
}

In this section, we compose an effective theory 
for the study of $^{13}$C($\alpha$,$n$)$^{16}$O at low energies. 
We study the reaction 
\bea
{}^{13}\textrm{C} + \alpha 
&\to& {}^{16}\textrm{O} + n\,,
\eea
where the energy difference  of the two channels, namely the $Q$ value, is
$Q = 2.2156~\textrm{MeV}$.\footnote{
When we set the zero energy at the ground state of $^{17}$O, 
the threshold energies of $n$-$^{16}$O and $\alpha$-$^{13}$C channels are
4.14~MeV and 6.359~MeV, respectively.
}
We chose a typical energy of the reaction for this study as 
the Gamow peak energy at $T\simeq 90 \times 10^6$~K 
in the low-mass AGB stars. It would be $E = 190\pm 40$~keV, 
where $E$ is the energy of the initial $\alpha$-$^{13}$C system 
in the center-of-mass frame.
The typical momentum scales would be 
${\cal Q} = \sqrt{2\mu_{\alpha C} E_G} = 33$~MeV where $\mu_{\alpha C}$ 
is the reduced mass of $\alpha$-$^{13}$C system, and we have fixed
the Gamow peak energy at $E_G = 0.19$~MeV.
Because the typical length scale, $\hbar  c/{\cal Q} = 6.0$~fm, is 
large compared to the sizes of the nuclei, 
we introduce the neutron and nuclei as 
point-like particles.

The spin, parity, 
and isospin states of the ground states of the nuclei and neutron are 
\bea
{}^{13}\textrm{C} (1/2^-;1/2)\,,
\alpha(0^+;0)\,,
{}^{16}\textrm{O} (0^+;0)\,,
n (1/2^+;1/2)\,,
\eea
and, in this work, we assume that the reaction occurs through the resonance 
states of $^{17}$O, which are effective around the energy of the initial 
$\alpha$-$^{13}$C state. 
There is no bound state of $^5$He, 
and the breakup energy of $^{13}$C in $n$-$^{12}$C channel 
is 4.94635~MeV 
and the energy of the first excited state of $^{13}$C
is 3.089~MeV.
They are significantly larger than the energy, $E=1$~MeV, 
at which we truncate the data set for parameter fit in this work.
Thus, $^5$He-$^{12}$C and $n$-$\alpha$-$^{12}$C channels 
are regarded as irrelevant degrees of freedom at high energies.
We may employ the two channels, 
\bea
\alpha + {}^{13}\textrm{C}, \ \ \ 
n + {}^{16}\textrm{O},
\eea
are relevant degrees of freedom 
at low energies and describe the reaction 
in terms of the resonant states of $^{17}$O. 

\begin{figure}
	\begin{center}
		\includegraphics[width=6cm]{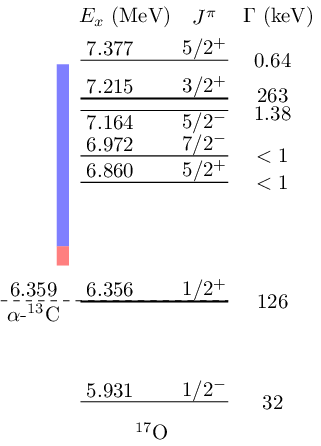}
		\caption{Level diagram of seven resonant states of $^{17}$O
		close to the $\alpha$-$^{13}$C breakup energy. 
		A vertical blue bar represents the energy range of the 
		experimental data to which we fit the parameters.
		A vertical red bar is that of the Gamow peak energy to 
		which the $S$ factor of $^{13}$C($\alpha$,$n$)$^{16}$O
		is extrapolated. 
		}
		\label{fig;levels}
	\end{center}
\end{figure}
%
In Fig.~\ref{fig;levels}, we display seven resonant states of $^{17}$O
around the $\alpha$-$^{13}$C breakup energy~\cite{twc-npa93}.
We also show the energy range of the experimental data of the $S$ factor
of $^{13}$C($\alpha$,$n$)$^{16}$O, by the blue band, at 
$E=0.23$~MeV to 1~MeV, and that of the Gamow-peak energy of 
the $\alpha$-$^{13}$C system, by the red band, at $E_G=0.190\pm 0.040$~MeV
in the figure. 
We include three states in this study; 
$3/2^+$ state at $E_x=7.215$~MeV ($\Gamma = 263$~keV) and $5/2^-$ state 
at $E_x = 7.164$~MeV ($\Gamma=1.38$~keV) in the energy range of the 
experimental data. The two $7/2^-$ and $5/2^+$ states are not included
because the widths are very small, $<1$~keV, and they do not appear in 
the experimental data. 
We include the $1/2^+$ state at $E_x=6.356$~MeV ($\Gamma = 126$~keV)
of a large width as a background contribution from low energy. 
While we do not include the largest and smallest states in the figure, 
the $5/2^+$ state at $E_x=7.377$~MeV and the $1/2^-$ state at $E_x=5.931$~MeV
because their widths may be small to give a significant contribution to 
the extrapolation of the $S$ factor to $E_G$. 
\begin{table}
	\begin{center}
	\begin{tabular}{|cccccc|}
		\hline
		$J^\pi$ & $E_x$ (MeV) & $\Gamma$ (keV) & $E$ (MeV) &
		$\alpha$-$^{13}$C & $n$-$^{16}$O \cr
		\hline 
		$1/2^+$ & 6.356(8) & 124(12) & $-$0.003(8) & 
		$l = 1$ & $l=0$ \cr 
		$5/2^-$ & 7.16586(17) & 1.38(5) & 0.80686(17)  & 
		$l=2$ & $l=3$ \cr
		$3/2^+$ & 7.202(10) & 280(30) & 0.843(10)  & 
		$l = 1$ & $l = 2$ \cr
		\hline
	\end{tabular}
	\caption{Energies and widths of the resonant states of $^{17}$O
		near the $\alpha$-$^{13}$C breakup threshold 
		energy~\cite{twc-npa93}.
		$E$ is the energy of the $\alpha$-$^{13}$C system 
		where $E=E_x - S_\alpha$;
		$S_\alpha$ is the 
		$\alpha$ separation energy from $^{17}$O is 
		$S_\alpha = 6.359$~MeV.
		In the fourth and fifth columns, the relative angular
		momenta between two clusters for $\alpha$-$^{13}$C
		and $n$-$^{16}$O channels are presented. 
	} \label{table;resonant_states}
	\end{center}
\end{table}
In Table \ref{table;resonant_states}, we summarize the three states and 
list the relative angular momenta between two clusters
of $\alpha$-$^{13}$C and $n$-$^{16}$O channels for the resonant 
states of $^{17}$O. 

When extrapolating the $S$ factor of $^{13}$C($\alpha$,$n$)$^{16}$O to $E_G$,
we presume that the energy dependence of the $S$ factor can be described
by the tails of the two broad resonance states at both low and high
energy sides of $E_G$, the $1/2^+$ state at $E=-0.003$~MeV ($E_x=6.356$~MeV)
and the $3/2^+$ state at $E=0.843$~MeV ($E_x=7.202$~MeV).
As discussed in Ref.~\cite{hfvk-epja21}, in the regions near the resonant 
energies, the experimental data are well described by two parameters,
the resonant energies and widths. 
While in the floor region between the two peaks,
long tails of broad resonances outside of the two peaks may contribute. 
Those contributions 
outside of the two peaks are taken into account
by the higher-order terms,
the shape parameters $a$ and $b$, 
of the two resonances. 

We have chosen the typical momentum scale as ${\cal Q}=33$~MeV 
for the reaction in the low-mass AGB stars above. 
Some large energy scales in the present study can be listed as
follows: 
the first resonant $1/2^+$ state of $^{13}$C, $E = 3.089$~MeV,
the breakup threshold energy of the $n$-$\alpha$-$^{12}$C channel,
$E = 4.95$~MeV, and a broad resonant $3/2^-$ state of $^{17}$O at
$E_x = 7.542(20)$~MeV and $\Gamma = 500(50)$~keV, which appears at
$E= 1.183$~MeV. So we may choose a high energy scale as $E_H= 1.2$~MeV,
and the large momentum scale is $\Lambda_H = \sqrt{2\mu_{\alpha C}E_H}
= 82$~MeV. Thus, we have the parameter of the perturbative
expansion as ${\cal Q}/\Lambda_H = 0.4$. 
We note that because momentum $\vec{p}$ is a vector,
and, 
the terms are expanded in terms of the energy $E$ in Eq.~(\ref{eq;LR}). 
Using the energy-momentum relation, $E=p^2/(2\mu_{\alpha C})$,
the corrections appear as a power of a scalar quantity, 
$({\cal Q}/\Lambda_H)^2 = 0.16$.~\footnote{
	In this study, we have very large quantities compared to the large 
	scale, $\Lambda_H = 82$~MeV, such as the reduced masses, 
$\mu_{\alpha C} = 2850~\textrm{MeV}$ and
$\mu_{n O} = 884~\textrm{MeV}$,
and the inverse of the Bohr radius of the $\alpha$-$^{13}$C system, 
$\kappa = Z_\alpha Z_{13C} \alpha_E \mu_{\alpha C} = 249.6$~MeV 
where $Z_\alpha$ and $Z_{13C}$ are the numbers of protons
in the nuclei and $\alpha_E$ is the fine structure constant. 
}

We encounter difficulty fitting the parameters of the $1/2^+$ state
because the data of the $S$ factor do not cover the resonant peak of the 
$1/2^+$ state. In general, we have five parameters for the $1/2^+$ state;
the resonant energy $E_R$, the neutron decay width $\Gamma_R^n$,
the $\alpha$ decay width $\Gamma_R^\alpha$ (equivalently the wavefunction
normalization factor $Z_R$), and two shape parameters
$a$ and $b$ in higher order. The energy range of the data is $E=0.23$~MeV
to 1~MeV, and at the lowest energy of the data, we may fit a shape
of the resonance tail, but it is not possible to adequately fit all five
parameters to the data.
To fit all of the five parameters to experimental data, we need to employ
another set of data, such as the cross section of elastic $n$-$^{16}$O
scattering. We postpone it to future work.
In this restricted situation, it is unavoidable to introduce a model fit. 

In the model fit for the present study, 
we fix the resonant energy as $E_R = -0.003$~MeV, 
and exclude the two shape constants $a$ and $b$ from the parameter fit.
Because $\Gamma_R^n$ and $\Gamma_R^\alpha$ equally contribute 
to the numerator of the reaction amplitude in Eq.~(\ref{eq;S1p}), 
we choose $\Gamma_R^n$ and $\Gamma_R^\alpha$
(equivalently $Z_R$) as two free parameters for the $1/2^+$ state. 

\vskip 2mm \noindent 
{\bf 3. Effective Lagrangian}

In this section, we construct the effective Lagrangian for the study of 
$^{13}$C($\alpha$,$n$)$^{16}$O at low energies. 
As discussed in the previous section, 
we have introduced the $\alpha$-$^{13}$C and $n$-$^{16}$O channels 
and the resonant $1/2^+$, $5/2^-$, $3/2^+$ states of $^{16}$O
in this study. 
The effective Lagrangian for the present study is presented as 
\bea
{\cal L} = {\cal L}_0 + {\cal L}_R +  {\cal L}_{int}\, ,
\eea
where ${\cal L}_0$ is the Lagrangian for 
$\alpha$, $^{13}$C, $n$, and $^{16}$O in the initial and final states,
${\cal L}_R$ is that for the resonant states of $^{17}$O, and 
${\cal L}_{int}$ is that for the interactions of the fields. 

The Lagrangian ${\cal L}_0$ is standard and one has
\bea
{\cal L}_0 &=& 
\phi_\alpha^* \left(
i \partial_0
+ \frac{1}{2m_\alpha}\nabla^2
\right) \phi_\alpha
+\psi_{13C}^\dagger \left(
i \partial_0
+ \frac{1}{2m_{13C}}\nabla^2
\right) \psi_{13C}
\nnb \\
&& 
+ \psi_n^\dagger \left(
i \partial_0
+ \frac{1}{2m_n}\nabla^2
\right) \psi_n
+ \phi_O^*\left(
i \partial_0
+ \frac{1}{2m_O} \nabla^2
\right) \phi_O \,,
\label{eq;L0}
\eea
where $\phi_\alpha$ and $\phi_O$ are non-relativistic scalar fields for 
$\alpha$ and $^{16}$O and $\psi_n$ and $\psi_{13C}$ are 
non-relativistic spin-1/2 fields for $n$ and $^{13}$C. 
The expressions of the inverse propagators of those fields 
are presented in Appendix C. 

The Lagrangian ${\cal L}_R$ for the resonant states of $^{17}$O
as non-relativistic fields is given as 
\bea
{\cal L}_R &=& 
\sum_{n=0}^N C_{1p}^{(n)} d_{1p}^\dagger 
\left(
i\partial_0 + \frac{1}{2m_{1p}}\nabla^2
\right)^n
d_{1p} 
+\sum_{n=0}^N C_{5m}^{(n)} d_{5m}^\dagger 
\left(
i\partial_0 + \frac{1}{2m_{5m}}\nabla^2
\right)^n
d_{5m} 
\nnb \\ && 
+ \sum_{n=0}^N C_{3p}^{(n)} d_{3p}^\dagger 
\left(
i\partial_0 + \frac{1}{2m_{3p}}\nabla^2
\right)^n
d_{3p}\,,
\label{eq;LR}
\eea
where $d_{1p}$, 
$d_{5m}$, 
$d_{3p}$, 
are the fields for the 
resonant 1/2$^+$, 
5/2$^-$, 3/2$^+$ 
states of $^{17}$O,
and $C_{1p}^{(n)}$, 
$C_{5m,j}^{(n)}$, $C_{3p}^{(n)}$ 
are the coupling 
constants to renormalize the infinities from self-energy terms, 
and the values of the constants are fixed by using the experimental data. 
$d_{1p}$ is the two-component spin-1/2 spinor,
$d_{5m}$ is the six-component spin-5/2 spinor, and 
$d_{3p}$ is the four-component spin-3/2 spinor. 
$m_{1p}$, $m_{5m}$, $m_{3p}$ 
are the masses corresponding to the
resonant states of $^{17}$O.
We note that, when one chooses the center-of-mass frame, 
the total three momenta of the fields vanish; 
one does not need to know the values of the masses of the resonant states. 
In the present work, 
we choose the number of the expansion terms, $N=1$, for the 
$1/2^+$ and $5/2^-$ states and $N=3$ for the $3/2^+$ state. 
The expressions of the inverse of the bare propagators are 
presented in Appendix C. 

The Lagrangian ${\cal L}_{int}$ to describe the interactions 
between the resonant states and the two cluster states is given as 
\bea
{\cal L}_{int} &=& 
-y_{1p}\left[
	d_{1p}^\dagger 
	\left(
	P^{(l=1)}_{1/2^+,i}
	\psi_{13C} 
	O^{(l=1)}_i
	\phi_\alpha\right)
	+ \left(
	P^{(l=1)}_{1/2^+,i}
	\psi_{13C} 
	O^{(l=1)}_i
	\phi_\alpha
	\right)^\dagger d_{1p}
	\right]
\nnb \\ && 
-y_{1p}'\left[
	d_{1p}^\dagger 
	\left(
	P^{(l=0)}_{1/2^+}
	\psi_n
	O^{(l=0)}
	\phi_O
	\right)
	+ \left(
	P^{(l=0)}_{1/2^+}
	\psi_n
	O^{(l=0)}
	\phi_O
	\right)^\dagger d_{1p}
	\right]
\nnb \\ && 
-y_{5m}\left[
	d^\dagger_{5m}
	\left(
	P_{5/2^-,ij}^{(l=2)}
	\psi_{13C}O^{(l=2)}_{ij}\phi_\alpha
	\right)
	+ \left(
	P_{5/2^-,ij}^{(l=2)}
	\psi_{13C}O^{(l=2)}_{ij}\phi_\alpha
	\right)^\dagger 
	d_{5m}
	\right]
\nnb \\ &&
-y_{5m}' \left[
	d^\dagger_{5m}
	\left(
	P_{5/2^-,ijk}^{(l=3)} 
	\psi_n O^{(l=3)}_{ijk} \phi_O
	\right)
	+ \left(
	P_{5/2^-,ijk}^{(l=3)} 
	\psi_n O^{(l=3)}_{ijk} \phi_O
	\right)^\dagger
	d_{5m}
	\right]
\nnb \\ &&
-y_{3p}\left[
	d_{3p}^\dagger 
	\left(
	P^{(l=1)}_{3/2^+,i}
	\psi_{13C} 
	O_i^{(l=1)}
	\phi_\alpha\right)
	+ \left(
	P^{(l=1)}_{3/2^+,i}
	\psi_{13C}
	O_i^{(l=1)}
	\phi_\alpha 
	\right)^\dagger d_{3p}
	\right]
\nnb \\ && 
-y_{3p}'\left[
	d_{3p}^\dagger 
	\left(
	P^{(l=2)}_{3/2^+,ij}
	\psi_{n}
	O^{(l=2)}_{ij}
	\phi_O\right)
	+ \left(
	P^{(l=2)}_{3/2^+,ij}
	\psi_n
	O^{(l=2)}_{ij}
	\phi_O
	\right)^\dagger d_{3p}
	\right]
\,,
\label{eq;Lint}
\eea
where 
$P^{(l=0)}_{1/2^+}$, 
$P^{(l=1)}_{1/2^+}$, 
$P^{(l=2)}_{5/2^-}$, 
$P^{(l=3)}_{5/2^-}$, 
$P^{(l=1)}_{3/2^+}$, 
$P^{(l=2)}_{3/2+}$ 
are the projection operators 
of the two cluster states 
to $J^\pi=1/2^+$, 
$J^\pi=5/2^-$, 
$J^\pi=3/2^+$ states 
with the relative angular momenta $l=0,1,2,3$.
$O^{(l=0)}$, 
$O^{(l=1)}_i$, 
$O^{(l=2)}_{ij}$, 
$O^{(l=3)}_{ijk}$, 
are the projection operators of the two cluster channels 
for $l=0,1,2,3$ states, 
respectively. 
The expressions of the projection operators of the relative angular
momenta, 
$O^{(l=0)}$, 
$O^{(l=1)}_i$, 
$O^{(l=2)}_{ij}$, 
$O^{(l=3)}_{ijk}$, 
are presented in Appendix B. 

For the $\alpha$-$^{13}$C channel, we have the projection operators,
$P^{(l=1)}_{1/2^+,i}$, $P^{(l=2)}_{5/2^-,ij}$, $P^{(l=1)}_{3/2^+,i}$, as 
\bea
P^{(l=1)}_{1/2^+,i} &=& 
-\frac{1}{\sqrt3}\sigma^T_i \,,
\ \ \ 
P^{(l=2)}_{5/2^-,ij} = \frac{1}{\sqrt5} R_i^T L_j^T\,,
\ \ \ 
P^{(l=1)}_{3/2^+,i} = S_i^T\,,
\label{eq;operators_aC}
\eea
where $\sigma_i$ are the Pauli matrices and $S_i$ are spin-1/2 to spin-3/2 
projection matrices~\cite{fhr-epja12}.
In addition, $(M_{ji}=) L_jR_i$ are spin-1/2 to spin-5/2 projection matrices.
For the $n$-$^{16}$O channel, we have the projection operators, 
$P_{1/2^+}^{(l=0)}$, 
$P_{5/2^-,ijk}^{(l=2)}$,
$P_{3/2^+,ij}^{(l=2)}$, as 
\bea
P^{(l=0)}_{1/2^+} &=& 1 \,,
\ \ \ 
P^{(l=3)}_{5/2^-,i} = -\frac{1}{\sqrt{15}}R^T_iL_j^T\sigma^T_k\,,
\ \ \ 
P^{(l=2)}_{3/2^+,ij} = 
- \sqrt{\frac25}S^T_i\sigma^T_j\,.
\label{eq;operators_nO}
\eea
In Appendix A, 
the expressions of the matrices, $S_i$, $L_j$, $R_k$, are presented
and the projection operators are discussed in detail.  

\vskip 2mm \noindent
{\bf 4.  $S$ factors and reaction amplitudes 
}

The $S$ factor of the $^{13}$C($\alpha$,$n$)$^{16}$O reaction in the center-of-mass frame is defined by 
\bea
S = \sigma(E) E  \exp(2\pi\eta)\,,
\eea
where $\sigma(E)$ is the total cross section 
and $\eta$ is the Sommerfeld parameter.
The total cross section is presented as 
\bea
\sigma(E) = 
\int 
d\Omega_{\hat{p}'}
\frac{d\sigma(E) }{
	d\Omega_{\hat{p}'}}
= \frac{\mu'\mu}{8\pi^2}\frac{p'}{p}
\int d\Omega_{\hat{p}'}
\sum_{spin}|A|^2\,,
\eea
where $\mu'=\mu_{nO}$ and $\mu = \mu_{\alpha 13C}$, and  
$\mu_{nO}$ and $\mu_{\alpha 13C}$ are the reduced masses of 
$n$-$^{16}$O and $\alpha$-$^{13}$C, respectively. 
$p'$ and $p$ are the magnitudes of the relative momenta of
the final $n$-$^{16}$O and initial $\alpha$-$^{13}$C states, respectively,
$p'=\sqrt{2\mu' (E+Q)}$ and $p=\sqrt{2\mu E}$, where $Q$ is the $Q$ value
of the reaction. 
$A$ is the reaction amplitude. 

\begin{figure}
\begin{center}
  \includegraphics[width=12cm]{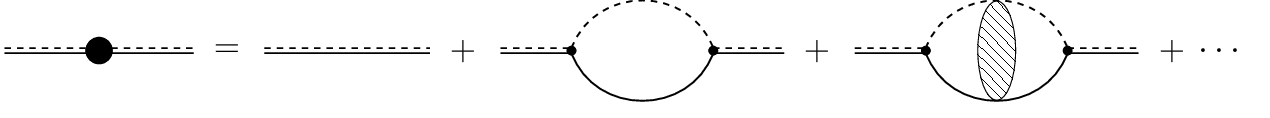}
\caption{
Diagrams for dressed $^{17}$O propagators.
	A thick and thin double-dashed line with and without a filled circle
represent a dressed and bare $^{17}$O propagators, respectively.
	A loop diagram with (without) a shaded ellipse represents
	the $n$-$^{16}$O ($\alpha$-$^{13}$C) self-energy term.
	A shaded ellipse represents the Coulomb Green's function. 
}
\label{fig;propagators}       
\end{center}
\end{figure}

\begin{figure}
\begin{center}
  \includegraphics[width=3cm]{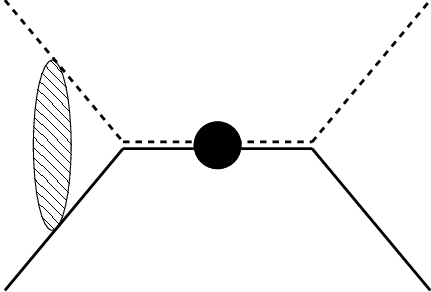}
\caption{
	Feynman diagram of the amplitudes of $^{13}$C($\alpha$,$n$)$^{16}$O 
	reaction. 
	A line (dashed line) in the initial state 
	represents a propagator of $^{13}$C ($\alpha$), and
	a line (dashed line) in the final state 
	represents a propagator of n ($^{16}$O).
	A shaded ellipse in the initial state represents 
	the Coulomb wavefunction. 
	See the caption in Fig.~\ref{fig;propagators} as well. 
}
\label{fig;amplitudes}       
\end{center}
\end{figure}

We decompose the $S$ factor as
\bea
S = S_{1p} + S_{5m} + S_{3p}\,,
\eea
where $S_{1p}$, $S_{5m}$, $S_{3p}$ are the $S$ factors obtained from 
the reaction amplitudes for the resonant $1/2^+$, $5/2^-$, $3/2^+$ states
of $^{17}$O, respectively. 
The reaction amplitudes, $A_{1p}$, $A_{5m}$, $A_{3p}$, 
for $S_{1p}$, $S_{5m}$, $S_{3p}$, respectively,  
are calculated from Feynman diagrams in Figs.~\ref{fig;propagators}
and \ref{fig;amplitudes}. 
In the following subsections, we construct the reaction amplitudes and 
$S$ factors for the resonant $1/2^+$, $5/2^-$, $3/2^+$ states of $^{17}$O,
while we display and discuss in detail the components of the amplitudes, 
vertex functions, 
initial Coulomb wavefunctions, self-energies, and dressed propagators 
of the resonant states of $^{17}$O in Appendix C. 

\vskip 2mm \noindent
{\bf 4.1 Reaction amplitude and $S$ factor for the $1/2^+$ resonant 
state of $^{17}$O
} 

We have the reaction amplitude for the $1/2^+$ resonant state of $^{17}$O 
as 
\bea
A_{1p} &=& 
-\Gamma_{dnO}^{(1p)} D_{1p}(p) \Psi_{d \alpha C}^{(1p)}(p)
\nnb \\ &=& 
- \frac{2\pi}{\sqrt3 \mu} 
\frac{ e^{i\sigma_1} \sqrt{p^2 + \kappa^2}C_\eta 
\chi_{n}^\dagger\vec{\sigma}^T\cdot \hat{p}\chi_C
	}{
		\sum_{n=0}^N \frac{2\pi}{y'y}C_{1p}^{(n)} E^n
		+ \frac{1}{3\mu} \left(\frac{y}{y'} 2\kappa H_1(p)
		\right))
		+ i \mu'\frac{y'}{y} p'
	}\,,
	\label{eq;A1p}
\eea
where $\Psi_{d\alpha C}^{(1p)}(p)$ is 
the 
vertex function 
of the initial $\alpha$-$^{13}$C state for $l=1$ 
including the initial Coulomb wavefunction to $d_{1p}$,  
$D_{1p}(p)$ is the dressed propagator of the resonant $1/2^+$ state 
of $^{17}$O, and $\Gamma_{dnO}^{(1p)}$ is the vertex function of $d_{1p}$ 
to the final $n$-$^{16}$O state for $l=0$. 
Those three components of the amplitude are presented in Appendix C. 
In addition, 
$\sigma_1$ is the Coulomb phase shift for $l=1$,
and the function $H_1(p)$ is obtained as\footnote{
	$H_l(p)$ functions are defined as~\cite{sa-prc23}
\bea
H_l(p) &=& W_l(p) H(\eta)\,,
\ \ \ 
W_l(p) = \left(\frac{\kappa^2}{l^2} + p^2 \right) W_{l-1}(p)\,,
\ \ \ W_0(p)=1\,.
\nnb 
\eea
}
\bea
H_1(p) &=& W_1(p)H(\eta) = (p^2+\kappa^2)H(\eta)\,,
\label{eq;H1}
\eea
with
\bea
H(\eta) = \psi(i\eta) + \frac{1}{2i\eta} - \log(i\eta)\,,
\eea
where 
$\eta$ is the Sommerfeld 
parameter, $\eta= \kappa/p$, $\kappa$ is the inverse of the Bohr radius
of the $\alpha$-$^{13}$C system, 
$\kappa = Z_\alpha Z_{13C} \mu\alpha_E$; 
$Z_\alpha$ and $Z_{13C}$ are the numbers of protons in the nuclei and
$\alpha_E$ is the fine structure constant, $\alpha_E=1/137.036$. 
$\psi(z)$ is the digamma function. 

Using a relation 
\bea
2\kappa ImH_1(p) = p(p^2+\kappa^2) C_\eta^2\,,
\label{eq;ImH1}
\eea
where
\bea
C_\eta^2 = \frac{2\pi\eta}{\exp(2\pi\eta) - 1}\,,
\eea
we define the $\alpha$ and neutron vertices
for the $^{13}$C($\alpha$,$n$)$^{16}$O reaction for
the resonant $1/2^+$ state of $^{17}$O as 
\bea
\Gamma^\alpha_{1p}(E) &=& 
2Z_{1p} p(p^2+\kappa^2)C_\eta^2\,,
\label{eq;Gamma_a}
\\
\Gamma^n_{1p}(E) &=& 6Z_{1p} \mu'\mu \frac{y'^2}{y^2}p'
= \Gamma^n_{R(1p)} \sqrt{
	\frac{E+Q}{-B+Q}}\,,
\eea
where $Z_{1p}$ is a wavefunction normalization factor and 
$\Gamma_{R(1p)}^n$ is the neutron width of the $1/2^+$ state at
$E=-B$. 
$Z_{1p}$ is obtained 
when the function, $2\kappa H_1(p)$, is normalized in the 
denominator of the amplitude as
\bea
\frac{6\pi\mu}{y^2}\sum C_{1p}^{(n)}E^n + 2\kappa Re H_1(p) 
= Z_{1p}^{-1}\left(E + B + \cdots\right)\,,
\label{eq;Z1p}
\eea
where the inverse of the propagator should vanish at $E=-B$, and
we neglect the higher-order terms, denoted by the dots, 
in the denominator of the amplitude for the $1/2^-$ state. 
The wavefunction normalization factor, $Z_{1p}$, 
is related to the asymptotic normalization coefficient (ANC)  of
the 1/2$^+$ state of $^{17}$O, 
as a bound state of the $\alpha$-$^{13}$C system,
by using a formula~\cite{irs-prc84}, 
\bea
|C_b|_{1p} = \gamma_{1p} \Gamma(2+\kappa/\gamma_{1p}) 
\sqrt{2\mu Z_{1p}}\,,
\label{eq;ANC}
\eea
where $\gamma_{1p}$ is the binding momentum of the $1/2^+$ state of $^{17}$O,
$\gamma_{1p}=\sqrt{2\mu B}$, and $\Gamma(z)$ is the gamma function.  

Thus, we have the reaction amplitude of the $1/2^+$ state of $^{17}$O as 
\bea
A_{1p} &=& - \frac{\pi}{\sqrt{\mu'\mu p'p}} 
\frac{e^{i\sigma_1} \sqrt{\Gamma^\alpha_{1p}(E) \Gamma^n_{1p}(E)}
}{
	E + B  
	+ i \frac12\left(
		\Gamma^\alpha_{1p}(E) + \Gamma^n_{1p}(E) 
	\right)
}
\chi_n^\dagger\vec{\sigma}^T\cdot\hat{p}\chi_C
\,,
\eea
and the $S$ factor of the $1/2^+$ resonant state of $^{17}$O is obtained as
\bea
S_{1p} = \frac{\pi}{2\mu}
\frac{\Gamma^\alpha_{1p}(E)\Gamma^n_{1p}(E)
e^{2\pi\eta}
}{
	(E+B)^2 + \frac14\left(
	\Gamma^\alpha_{1p}(E) + \Gamma^n_{1p}(E)
	\right)^2
} 
\,.
\label{eq;S1p}
\eea
The $S_{1p}$ factor has three parameters, $B$, $Z_{1p}$, and 
$\Gamma_{R(1p)}^n$. 
In this work, we fix $B=3$~keV and two 
parameters, $Z_{1p}$ and $\Gamma^n_{R(1p)}$, are fitted to the experimental
data. 

\vskip 2mm \noindent
{\bf 4.2 Reaction amplitude and $S$ factor for $5/2^-$ state of $^{17}$O} 

The reaction amplitude of $^{13}$C($\alpha$,$n$)$^{16}$O 
for the resonant $5/2^-$ state of $^{17}$O is
\bea
A_{5m} &=& - \Gamma_{dnO}^{(5m)}(p')D_{5m}(p) \Psi_{d\alpha C}^{(5m)}(p) 
\nnb \\ &=&
- \frac{1}{10\sqrt3}
\frac{y'y}{\mu'^3\mu^2} 
\frac{
	\chi_n^\dagger
	\vec{\sigma}^*\cdot\vec{p}'\vec{L}^*\cdot\vec{p}'\vec{R}^*\cdot\vec{p}'
	\vec{R}^T\cdot \vec{p}\vec{L}^T\cdot\vec{p}
	\chi_C
	e^{i\sigma_2}\sqrt{(1+\eta^2)(4+\eta^2)}C_\eta
	}{
		\sum_{n=0}^N C_{5m}^{(n)} E^n
		+ \frac{2y^2}{15\mu^4} \left(
		\frac{\mu}{2\pi}2\kappa H_2(p)
		\right)
		+ i \frac{2y'^2}{45\mu'^6}\frac{\mu'}{2\pi}p'^7
	}
	\,,
	\label{eq;A5m}
\eea
where $\Psi_{d\alpha C}^{(5m)}(p)$ is the vertex function 
of the $\alpha$-$^{13}$C state for $l=2$ 
including the initial Coulomb wavefunction
to $d_{5m}$,  
$D_{5m}(p)$ is the dressed propagator of the $5/2^-$ state of $^{17}$O,
and $\Gamma_{dnO}^{(5m)}(p')$ is the vertex function of $d_{(5m)}$
to the $n$-$^{16}$O state for $l=3$. 
Those three components of the amplitude
are presented in Appendix C. 
$\sigma_2$ is the Coulomb phase shift for $l=2$, and 
the function $H_2(p)$ is given as
\bea
H_2(p) = W_2(p)H(\eta) = \frac14p^4(1+\eta^2)(4+\eta^2)H(\eta)\,.
\eea

Using a relation,
\bea
2\kappa Im H_2(p) = \frac14 p(p^2+\kappa^2) (4p^2+\kappa^2)C_\eta^2\,, 
\eea
and a similar relation to Eq.~(\ref{eq;Z1p}) to define $Z_{5m}$,
where $Z_{5m}$ is the wavefunction normalization factor,
we define the $\alpha$ and neutron vertices as 
\bea
\Gamma^\alpha_{5m}(E)  &=& 2 Z_{5m}
p\, W_2(p) C_\eta^2 
= \Gamma^\alpha_{R(5m)} 
\frac{pW_2(p)C_\eta^2}
     {p_rW_2(p_r)C_{\eta_r}^2}
\,,
\\
\Gamma^n_{5m}(E) &=& 
Z_{5m}  
\frac{2\mu^3y'^2}{3\mu'^5y^2} 
p'^7
= \Gamma^n_{R(5m)} \left(
\frac{E+Q}{E_{R(5m)} + Q}
\right)^{7/2}
\,,
\eea
where $p_r$ and $\eta_r=\kappa/p_r$ are the momentum and Sommerfeld factor
at $E=E_{R(5m)}$, $p_r = \sqrt{2\mu E_{R(5m)}}$; 
$E_{R(5m)}$ is the resonant energy of the $5/2^-$ state of $^{17}$O.
$\Gamma^\alpha_{R(5m)}$ and $\Gamma^n_{R(5m)}$ are the $\alpha$ and 
neutron widths of the $5/2^-$ state at $E=E_{R(5m)}$. 
We note that because 
$\Gamma^\alpha_{R(5m)}$ and $\Gamma^n_{R(5m)}$ are fitted to the 
experimental data, the factor $Z_{5m}$ can be an arbitrary constant. 
Thus, we have the reaction amplitude of the $^{13}$C($\alpha$,$n$)$^{16}$O
reaction of the resonant $5/2^-$ state of $^{17}$O as 
\bea
A_{5m} &=& - \frac{3\pi}{2\sqrt{\mu'\mu p'p}}
\frac{
	\chi_n^\dagger
	\vec{\sigma}^*\cdot\hat{p}'\vec{L}^*\cdot\hat{p}'\vec{R}^*\cdot\hat{p}'
	\vec{R}^T\cdot \hat{p}\vec{L}^T\cdot\hat{p}
	\chi_C
	e^{i\sigma_2}\sqrt{\Gamma^n_{5m}(E)\Gamma^\alpha_{5m}(E)}
	}{
		E - E_{R(5m)} 
		+ i \frac12\left(
		\Gamma^n_{5m}(E) + \Gamma^\alpha_{5m}(E) 
		\right)
	}\,.
\eea

To calculate the squared amplitude and angle integral in the phase space, 
we have
\bea
\int\frac{d\Omega_{\hat{p}'}}{4\pi}
\textrm{Tr}\left(
\vec{R}^\dagger\cdot\hat{p}' \vec{L}^\dagger \cdot \hat{p}' 
\vec{\sigma}^\dagger\cdot \hat{p}'\vec{\sigma}\cdot \hat{p}'
\vec{L}\cdot\hat{p}' \vec{R}\cdot \hat{p}'
\vec{R}^\dagger \cdot \hat{p} \vec{L}^\dagger \cdot \hat{p}
\vec{L}\cdot \hat{p} \vec{R}\cdot \hat{p} 
\right)^* = \frac83\,, 
\eea
where we have chosen $\hat{p}=(0,0,1)$.   
Thus, we have the $S$ factor of the $5/2^-$ state of $^{17}$O as
\bea
S_{5m} = \frac{3\pi}{2\mu}
\frac{\Gamma^n_{5m}(E) \Gamma^\alpha_{5m}(E)e^{2\pi\eta}}{
	(E - E_{R(5m)})^2 + \frac14\left(
	\Gamma^n_{5m}(E) + \Gamma^\alpha_{5m}(E) 
	\right)^2
}\,.
\eea
The $S_{5m}$ factor has three parameters,
$E_{R(5m)}$, $\Gamma^n_{R(5m)}$, $\Gamma^\alpha_{R(5m)}$.
Those parameters are fitted to the experimental data. 

\vskip 2mm \noindent
{\bf 4.3 Reaction amplitude and $S$ factor for $3/2^+$ state of $^{17}$O} 

The reaction amplitude of $^{13}$C($\alpha$,$n$)$^{16}$O 
for the resonant $3/2^+$ state of $^{17}$O is
\bea
A_{3p} &=& - \Gamma_{dnO}^{(3p)}(p') D_{3p}(p) \Psi_{d\alpha C}^{(3p)}(p)
\nnb \\ &=& 
- \sqrt{\frac25} \frac{y'y}{\mu'^2\mu}
\frac{\chi_n^\dagger \vec{\sigma}^*\cdot \vec{p}'\vec{S}^*\cdot\vec{p}'
\vec{S}^T\cdot\vec{p}\chi_C e^{i\sigma_1}\sqrt{1+\eta^2}C_\eta
	}{
		\sum_{n=0}^N C_{3p}^{(n)} E^n 
		+ \frac{y^2}{3\mu^2}\frac{\mu}{2\pi} (2\kappa H_1(p))
		+ \frac{2y'^2}{15\mu'^4}\frac{\mu'}{2\pi}(ip'^5) 
	}
	\,,
\eea
where $\Psi_{d\alpha C}^{(3p)}(p)$ is the vertex function 
of the $\alpha$-$^{13}$C state for $l=1$ to $d_{3p}$ 
including the initial Coulomb wavefunction,
$D_{3p}(p)$ is the dressed propagator of the resonant $3/2^+$ state of 
$^{17}$O, and $\Gamma^{(3p)}_{dnO}(p')$ is the vertex function of $d_{3p}$
to $n$-$^{16}$O state for $l=2$. 
Those three components of the reaction amplitude are presented in Appendix C. 

Using the relation in Eq.~(\ref{eq;ImH1}),  we define the $\alpha$ and 
neutron vertices of the $3/2^+$ state as 
\bea
\Gamma^\alpha_{3p}(E) &=& 
2Z_{3p} pW_1(p)C_\eta^2
= \Gamma^\alpha_{R(3p)} \frac{pW_1(p)C_\eta^2}{
	p_rW_1(p_r)C_{\eta_r}^2
}
\,,
\\
\Gamma^n_{3p}(E) &=& 
Z_{3p} \frac{2\mu y'^2}{5\mu'^3y^2} p'^5
= \Gamma^n_{R(3p)} \left(
\frac{E+Q}{E_{R(3p)} + Q}
\right)^{5/2}
\,,
\eea
where $Z_{3p}$ is the wavefunction normalization constant, 
$\Gamma^\alpha_{R(3p)}$ and $\Gamma^n_{R(3p)}$ are the 
$\alpha$ and neutron widths, and $p_r$ and $\eta_r$ are the momentum 
and Sommerfeld parameter at $E=E_{R(3p)}$.
$E_{R(3p)}$ is the resonant energy of the $3/2^+$ state of $^{17}$O. 
Because $\Gamma^\alpha_{R(3p)}$ and $\Gamma^n_{R(3p)}$ are fitted to
the experimental data, $Z_{3p}$ can be an arbitrary constant. 
Thus, we have the reaction amplitude of the $3/2^+$ state as 
\bea
A_{3p} &=& - \frac{3\pi}{\sqrt{\mu'\mu p'p}}
\frac{e^{i\sigma_1} 
\chi_n^\dagger \vec{\sigma}^*\cdot\hat{p}' \vec{S}^*\cdot\hat{p}'
\vec{S}^T\cdot \hat{p}\chi_C
\sqrt{\Gamma^n_{3p}(E)\Gamma^\alpha_{3p}(E)}
	}{
		E - E_{R(3p)} + R_{3p}(E)
		+ i \frac12\left(
		\Gamma^\alpha_{3p}(E) + \Gamma^n_{3p}(E) 
		\right)
	}\,,
\eea
where
\bea
R_{3p}(E) &=& a_{3p}(E-E_{R(3p)})^2 + b_{3p}(E-E_{R(3p)})^3 
\,.
\eea
The terms in $R_{3p}(E)$ are the higher order terms of the effective range
parameters, obtained by Fourier expansion around $E=E_{R(3p)}$. 
The coefficients, $a_{3p}$ and $b_{3p}$,
describe the shape of the resonant peak and are fitted to the experimental data. 
To calculate the squared amplitude and angle integral in the phase space, 
we have 
\bea
\int \frac{d\Omega_{\hat{p}'}}{4\pi} 
\textrm{Tr} \left(
\vec{S}\cdot\hat{p}\vec{S}^\dagger\cdot \hat{p}'
\vec{\sigma}^\dagger\cdot \hat{p}'
\vec{\sigma}\cdot\hat{p}'\vec{S}\cdot \hat{p}'\vec{S}^\dagger\cdot \hat{p}
\right)^*
&=& 
\frac49
\,,
\eea
where we have chosen $\hat{p}=(0,0,1)$, and
we have the expression of the $S$ factor of $^{13}$C($\alpha$,$n$)$^{16}$O
for the $3/2^+$ state of $^{17}$O as 
\bea
S_{3p} = \frac{\pi}{\mu}
\frac{\Gamma^n_{3p}(E) \Gamma^\alpha_{3p}(E) e^{2\pi\eta}}{
	\left(E - E_{R(3p)} + R_{3p}(E) \right)^2 
	+ \frac14\left(
	\Gamma^n_{3p}(E) + \Gamma^\alpha_{3p}(E) 
	\right)^2
}\,.
\eea
The $S_{3p}$ factor has five parameters, \{$E_{R(3p)}$, $\Gamma^n_{R(3p)}$,
$\Gamma^\alpha_{R(3p)}$, $a_{3p}$, $b_{3p}$\}, and they are fitted to
the experimental data. 

\vskip 2mm \noindent
{\bf 5. Numerical results} 

In the previous section, we have constructed the $S$ factors 
of $^{13}$C($\alpha$,$n$)$^{16}$O for the resonant $1/2^+$,
$5/2^-$, $3/2^+$ states of $^{17}$O in the cluster EFT.  
The $S$ factors contain eleven parameters 
while we fix $B$ as $B=3$~keV, 
and the remaining ten parameters, 
\{$\Gamma^n_{R(1p)}$, $Z_{1p}$, 
$E_{R(5m)}$, $\Gamma^n_{R(5m)}$, $\Gamma^\alpha_{R(5m)}$,
$E_{R(3p)}$, $\Gamma^n_{R(3p)}$, $\Gamma^\alpha_{R(3p)}$, $a_{3p}$, $b_{3p}$\},
are fitted to the experimental data. 
We employ the data sets of the $S$ factor 
from Bair and Haas (1973)~\cite{bh-prc73},
Drotleff et al. (1993)~\cite{detal-aj93},
Heil et al. (2008)~\cite{hetal-prc08},
LUNA collaboration (2021)~\cite{LUNA},
JUNA collaboration (2022)~\cite{JUNA}, 
and the data for the parameter fit are truncated at 
the energy $E=1$~MeV, 
just below the sharp resonant $5/2^+$ state of $^{17}$O.
Thus, the energy range of the parameter fit is $E= 0.230$ to 1~MeV, 
where the data at the smallest energy were reported from the LUNA collaboration. 
We perform a $\chi^2$ fit in an MCMC analysis by employing the {\tt emcee} 
package~\cite{fetal-pasp13}. By using the fitted values of the parameters,
we extrapolate the S factor to $E_G=0.19$~MeV and estimate its error in the 
MCMC analysis. 

\begin{figure}
	\centering
  \includegraphics[width=0.8\textwidth]{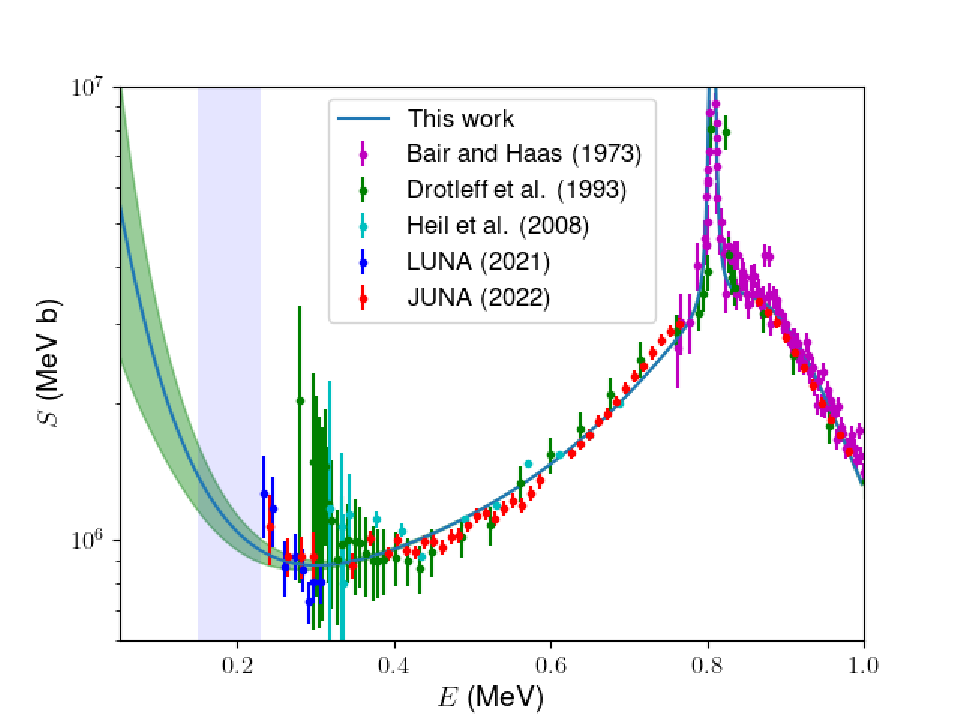}
	\caption{
		S factor of $^{13}$C($\alpha$,$n$)$^{16}$O 
		as a function of energy, $E$, of the initial
	$\alpha$-$^{13}$C state in the center-of-mass frame.
	A line is 
	plotted by using fitted values 
	of ten parameters to experimental data,  
	presented in the second column in 
	Table~\ref{table;fitted_parameters}.
	A band of the plotted line is obtained 
	by 16 to 84~\% samples of the MCMC analysis. 
	The experimental data are displayed in the figure as well. 
	A vertical band represents a range of the Gamow peak energy
	in the low-mass AGB stars. 
}
	\label{fig;S-factor10we}
\end{figure}

Because the resonant energy of the $1/2^+$ state is not covered by the
experimental data for the parameter fit, as mentioned before, 
we fix the resonant energy of the $1/2^+$ state as
$E=-B$ where $B=3$~keV. 
While the reported neutron width, 
$\Gamma^n_{R(1p)}=124(12)$~keV,
and the ANC of the $1/2^+$ state of $^{17}$O 
have significant uncertainties, and we do not fix them by using the 
reported values, but fit the tail from the low-energy pole to the data.  
We perform two runs of the parameter fit, 
a ten-parameter fit and a seven-parameter fit. 
In the first ten-parameter fit, we include 
all the data sets mentioned above 
but find a significantly large $\chi^2$ value, $\chi^2/N=8.20$, 
where $N$ is the number of data points, $N=216$. 
In addition, the recently reported data from the JUNA collaboration do not fit well. (See Fig.~\ref{fig;S-factor10we}.)
In the second seven-parameter fit, we include only the recent two data sets from the LUNA and JUNA collaborations. 
Because the two data sets do not have the data for the sharp resonant
$5/2^-$ state of $^{17}$O, we fix the three parameters, 
$E_{R(5m)}$, $\Gamma^n_{R(5m)}$, $\Gamma^\alpha_{R(5m)}$, 
by using the values fitted in the first ten parameter fit. 
The seven parameters 
\{$\Gamma^n_{R(1p)}$, $Z_{1p}$, 
$E_{R(3p)}$, $\Gamma^n_{R(3p)}$, 
$\Gamma^\alpha_{R(3p)}$, $a_{3p}$, $b_{3p}$\}
are fitted to the data.  

\begin{table}
\begin{center}
\begin{tabular}{c|cc|c}
\hline
	& 10 parameter fit & 7 parameter fit & Ref.~\cite{twc-npa93}  \cr
\hline
	$\Gamma^n_{R(1p)}$ (keV) & 147$^{+158}_{-83}$ & 136$^{+192}_{-100}$ &
	124(12)$^{*}$ \cr
	$Z_{1p}$ (MeV$^{-2}$) & 3.0$^{+3.3}_{-1.1}\times 10^{-3}$ & 
	1.5$^{+2.2}_{-0.9}\times 10^{-3}$\cr
	$E_{R(5m)}$ (MeV) & 0.806612(17) & $-$ & 0.80686(17) \cr
	$\Gamma^n_{R(5m)}$ (keV) & 2.98(4) & $-$ & 1.38(5)$^{*}$ \cr
$\Gamma^\alpha_{R(5m)}$ (eV) & 6.32(6) & $-$ \cr
	$E_{R(3p)}$ (MeV) & 0.8559(8) &  0.8416(15) & 0.843(10) \cr
	$\Gamma^n_{R(3p)}$ (keV) & 285(4) & 291(6) & 263(7)$^{*}$ \cr
	$\Gamma^\alpha_{R(3p)}$ (eV) & 99.8(11) & 85.8(17) \cr
	$a_{3p}$ (MeV) & 1.78(4) & 1.32(7) \cr
	$b_{3p}$ (MeV$^2$) & 1.52(13) & 0.51$^{+0.17}_{-0.14}$ \cr 
	\hline
	$\chi^2/N$ ($N$) & 8.20 (216) & 0.67 (56) \cr
	$|C_b|_{1p}$ (fm$^{-1/2}$) & 3.16$^{+3.48}_{-1.16}\times 10^{90}$ & 
	2.23$^{+3.27}_{-1.34}\times 10^{90}$ & 5.44$\times {10^{90}}^{**}$
	\cr
	$S$(0.19~MeV) (MeV b) & 
	$1.09(11)\times 10^6$ & $1.12(8)\times 10^6$ \cr
	\hline 
\end{tabular}
	\caption{
		Values of the parameters fitted to the experimental data of the 
	$S$ factor of $^{13}$C($\alpha$,$n$)$^{16}$O below $E=1$~MeV.
	In the second column, the values of ten parameters fitted 
	to the available experimental data 
	are displayed. 
	In the third column, those of the seven parameters 
	fitted to the data reported from the LUNA and JUNA collaborations 
	are presented. 
	In the last column, the values of resonant energies and widths in Ref.~\cite{twc-npa93} are displayed. 
	In the third row from the bottom, values of $\chi^2/N$ for the fit 
	where $N$ is the number of data, in the second row from the bottom,
	those of the ANC of $1/2^+$ state of $^{17}$O
	calculated using the fitted values of $Z_{1p}$, and in the 
	last row, those of $S$ factor at $E=0.19$~MeV are displayed. 
	See the text as well. 
	$^{*}$Total widths, $\Gamma^{tot}\simeq \Gamma^n_R$.  
	$^{**}$A value found, e.g., in Table II in Ref.~\cite{LUNA}. 
}
	\label{table;fitted_parameters}
\end{center}
\end{table}
In Table~\ref{table;fitted_parameters}, the fitted values of the 
parameters in the MCMC analysis are displayed. 
In the second column of the table, the 
fitted values of the ten-parameter fit, and in the third column, the 
fitted values of the seven-parameter fit are presented. 
In the last column, reference values in the literature are displayed. 
In the third row from the bottom, the $\chi^2/N$ values of the parameter fit
are presented, where $N$ is the number of data points.    
In the second row from the bottom, the values of the ANC of 
the $1/2^+$ state of $^{17}$O at $E=-2.69$~keV~\cite{detal-prc20} 
calculated by using the fitted values of $Z_{1p}$ 
are displayed. 
In the last column, the values and errors of the $S$ 
factor of $^{13}$C($\alpha$,$n$)$^{16}$O at $E=0.19$~MeV are presented. 

As seen in the table, the fitted values of $\Gamma^n_{R(1p)}$ and $Z_{1p}$ 
have about 100\% error bars, mainly because the experimental data 
do not cover the energy range for the $1/2^+$ state of $^{17}$O. 
The fitted values of $Z_{1p}$ are converted to the values of the ANC 
by using Eq.~(\ref{eq;ANC}), displayed in the second row from the 
bottom of the table. 
We find that the fitted values of 
$\Gamma^n_{R(1p)}$ and $|C_b|_{1p}$ agree with the values in the literature
within the large error bars. 
The fitted value of $E_{R(5m)}$ agrees well that in Ref.~\cite{twc-npa93},
while that of $\Gamma^n_{R(5m)}$ is about two times larger.
For the parameters for the $3/2^+$ state, two fitted values of $E_{R(3p)}$ 
agree with that of Ref.~\cite{twc-npa93} within the error bars, while 
two fitted values of $\Gamma^n_{R(3p)}$ are 8.3\% and 10.6\% larger than
that in Ref.~\cite{twc-npa93}.

\begin{figure}
	\centering
  \includegraphics[width=0.8\textwidth]{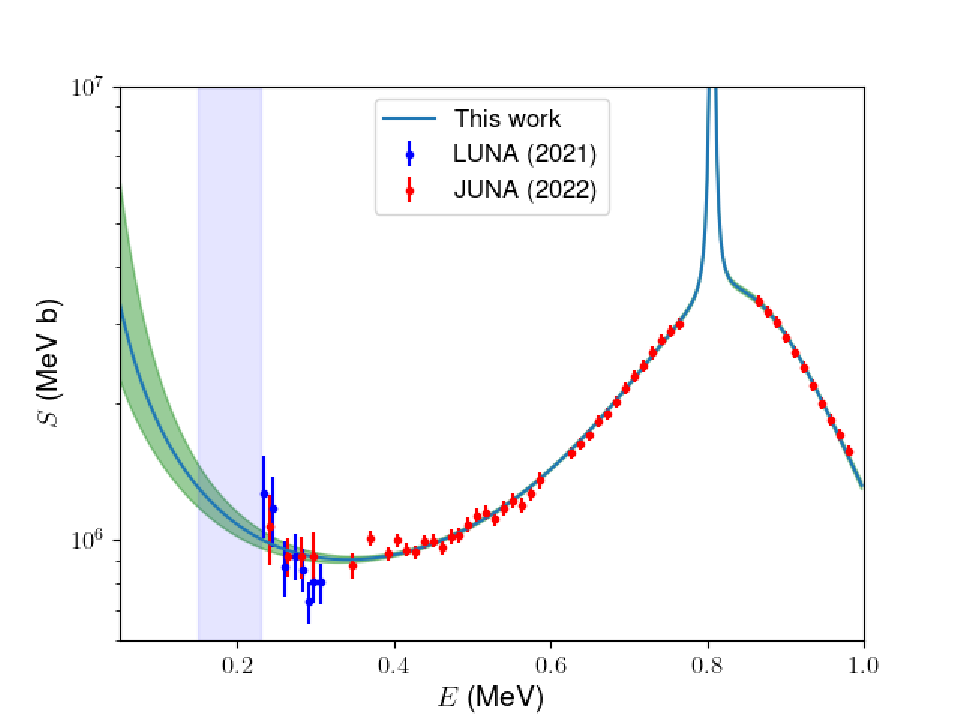}
	\caption{
		S factor of $^{13}$C($\alpha$,$n$)$^{16}$O 
		as a function of energy, $E$, of the initial
	$\alpha$-$^{13}$C state in the center-of-mass frame.
	A line is 
	plotted by using fitted values 
	of seven parameters to experimental data 
	of LUNA and JUNA collaborations, 
	presented in the third column in 
	Table~\ref{table;fitted_parameters}.
	See the caption of Fig.~\ref{fig;S-factor10we} as well. 
	}
	\label{fig;S-factor7we}
\end{figure}

In Figs.~\ref{fig;S-factor10we} and \ref{fig;S-factor7we}, 
we plot the $S$ factor of $^{13}$C($\alpha$,$n$)$^{16}$O as a function
of the energy $E$ of the initial $\alpha$-$^{13}$C system in the 
center-of-mass frame by using the two sets of fitted values 
of the parameters in Table~\ref{table;fitted_parameters}. 
The green bands of the fitted lines are obtained 
by 16 to 84~\% samples of the MCMC analysis. 
The experimental data for the parameter fit are also displayed in the figures. 
As presented in the table, the $\chi^2$ values of the parameter fit,
$\chi^2/N$ ($N$) are 8.20 (216) and 0.67 (56) for the ten and seven 
parameter fits, respectively. As mentioned above, we have a large $\chi^2$
value in the ten-parameter fit, and one can see that the recent data from JUNA
are not well fitted in Fig.~\ref{fig;S-factor10we}. 
In Fig.~\ref{fig;S-factor7we}, one can see that the data from LUNA and JUNA
are fitted well. 
Because we do not fix the two parameters, $\Gamma^n_{R(1p)}$ 
and $Z_{1p}$, of the $1/2^+$ state by using the values in the literature, 
the error bands of the $S$ factor increase 
as the energy $E$ decreases in the vertical band 
which represents a range of the Gamow peak energy in the low mass AGB stars. 
In the last row in Table~\ref{table;fitted_parameters}, we present 
our estimate of the $S$ factor at $E_G=0.19$~MeV. We find that both
values of the $S$ factor are in good agreement within the error bars, 
and we have 10.1\% and 
7.0\% errors in the $S$ factor at $E_G$. 
We note that the estimated errors are generated in the MCMC analysis,
and they reflect only the statistical errors from the data of the $S$ factor.
In the present study, we have employed a model fit for the parameters of the 
$1/2^+$ state, and the error estimate due to the perturbative expansion 
scheme is not available. 

\vskip 2mm \noindent
{\bf 6. Results and discussion } 

In the present study, we constructed an EFT 
for the $^{13}$C($\alpha$,$n$)$^{16}$O reaction at low energies. 
We included the resonant $1/2^+$, $5/2^-$, $3/2^+$ states of $^{17}$O
and the two open channels, $\alpha$-$^{13}$C and $n$-$^{16}$O states,
and wrote down an effective Lagrangian for the study. 
The expressions of the $S$ factors and reaction amplitudes of 
$^{13}$C($\alpha$,$n$)$^{16}$O were calculated.
We have eleven parameters in the amplitudes, 
and one of them was fixed by using the resonant energy of the $1/2^+$ state. 
The remaining parameters were fitted to the data of the $S$ factor 
at the energies below $E=1$~MeV. 
We performed two runs of the parameter fit by 
employing different sets of the experimental data.
By using the fitted values of the parameters, the $S$ factor was 
extrapolated to $E_G=0.19$~MeV, and we found that the uncertainties of the $S$ 
factor are about 10\% at $E_G$.  
As seen in Figs.~\ref{fig;S-factor10we} and \ref{fig;S-factor7we}, the error bands increase as the energy decreases 
in the range of the Gamow peak energy in the low-mass AGB stars
because the parameters,
$\Gamma^n_{R(1p)}$ and $Z_{1p}$, are not fitted well to the experimental 
data. 
We confirm, as pointed out in Ref.~\cite{detal-prc20},  
that the uncertainty of the $S$ factor mainly stems from 
the near-threshold state of $^{17}$O. 

The expressions of the $S$ factors of 
$^{13}$C($\alpha$,$n$)$^{16}$O obtained in the cluster EFT
are almost the same as those in the Breit-Wigner formula. 
See, e.g., Eq.~(2) in Ref.~\cite{getal-aj12}. 
As pointed out in Ref.~\cite{hbp-prc14} for the study of $^3$He($d$,$n$)$^4$He,
on one hand, the expression of the reaction amplitude calculated in an EFT can be obtained 
in the limit that the channel radius $a$ sends to zero in an
$R$-matrix analysis. In the present work, on the other hand, 
we introduced the higher order terms, $a_{3p}$ and $b_{3p}$, in the 
effective range expansion in the reaction amplitude 
for the $3/2^+$ state of $^{17}$O.  
They played a major role in fitting the shape of the $S$ factor 
at the low-energy side of the resonant $3/2^+$ state
without introducing a resonant state at high energy. 
Some extension in the traditional approach is possible 
within a framework of EFT. 

It was reported that even a factor of $\simeq 4$ uncertainty of 
$^{13}$C($\alpha$,$n$)$^{16}$O reaction at $T=90\times
10^6$~K gives a negligible influence on models 
of low-mass AGB stars~\cite{csg-npa05}. 
So the about 10\% uncertainty of the $S$ factor at $E_G$ obtained 
in this work would be a good estimate of 
the $^{13}$C($\alpha$,$n$)$^{16}$O reaction at low energies. 
While it may also be important to improve the uncertainty of the reaction
itself. 
The values of $\Gamma^n_{R(1p)}$ and $Z_{1p}$,
equivalently the ANC of the $1/2^+$ state of $^{17}$O, 
can be found in the literature, 
(we listed them in the tables.)
and, as mentioned, 
we fixed the resonant energy of the $1/2^+$ state as $B=3$~keV.
The resonant $1/2^+$ state has a relatively large width,
$\Gamma_{R(3p)}^n = 124(12)$~keV, and 
the reported value of the resonant energy of the $1/2^+$ state 
has a relatively large error bar, 
$B = 3\pm 8$~keV.
Moreover, the ANC is sensitive to the value of $B$.  
Thus, we postponed the study of the parameter sets for the $1/2^+$ state of
$^{17}$O to future work. 
To introduce other experimental data, such as  elastic $n$-$^{16}$O scattering,
in the study may be necessary to fix 
the parameters of the 1/2$^+$ state of $^{16}$O 
and reduce the errors of the $S$ factor of $^{13}$C($\alpha$,$n$)$^{16}$O 
at $E_G$ obtained in this work. 

\vskip 2mm \noindent
{\bf Acknowledgements}

The author would like to thank Gurgan Adamian for his suggestion to 
apply a cluster EFT to the study of nuclear reactions,
Chang Ho Hyun and Ki-Seok Choi 
for discussions in the early stage of the work,
and Jubin Park for discussions about the $R$-matrix analysis. 
This work was supported by the National Research Foundation grant
funded by the Korean government Ministry of Science and ICT (Grants
No. 2023R1A2C1003177 and No. RS-2025-16065411).

\vskip 2mm \noindent
{\bf Appendix A: Projection operators}

In this appendix, we present the expressions of the matrices,
$S_i$, $L_i$, $R_i$, and discuss the projection operators 
in Eqs.~(\ref{eq;operators_aC}) and (\ref{eq;operators_nO}). 
The expression of the operator $P_{1/2}^{(l=0)}$ is obvious,
$P_{1/2}^{(l=0)}=1$, for the $n$-$^{16}$O channel with $l=0$ and 
spin-1/2 states. 

\vskip 2mm
\noindent
$\bullet$  
$P_{1/2,i}^{(l=1)}$;  
$J^\pi=1/2^+$ state 
for the $\alpha$-$^{13}$C channel 
with $l=1$ and spin-1/2 states

For the term with the $y_{1p}$ coefficient in Eq.~(\ref{eq;y1p}),
in the center-of-mass frame, 
we choose the momentum of light particle $\alpha$ ($n$) as $+\vec{p}$
and that of heavy particle $^{13}$C ($^{16}$O) as $-\vec{p}$, and 
the projection operator, $O^{(l=1)}_i$, 
leads to $p_i/\mu_{\alpha C}$ where $\mu_{\alpha O}$ is the reduced mass
of $\alpha$ and $^{13}$C
($p_i/\mu_{nO}$ for the $n$-$^{16}$O channel
where $\mu_{nO}$ is the reduced mass of $n$ and $^{16}$O).
Thus one has
\bea
d_{1p}^\dagger\left(P_{1/2,i}^{(l=1)}\psi_{13C}O^{(l=1)}_i\phi_\alpha\right)
\to \frac{1}{\mu_{\alpha C}} 
\chi_{1p}^\dagger
\left(
- \frac{1}{\sqrt3}\vec{\sigma}^T\cdot \vec{p}
\right)
\chi_{C}\,,
\eea
where $\chi_{1p}$ and $\chi_{C}$ are the two-component spinors of
the $1/2^+$ state of $^{17}$O and $^{13}$C, respectively.  
Using the relations of the Cartesian tensors~\cite{sakurai},
\bea
\hat{p}_\pm = \mp \frac{1}{\sqrt2}(\hat{p}_1 \pm i \hat{p}_2) \,,
\ \ \
\sigma_\pm = \frac12(\sigma_1 \pm i \sigma_2)\,,
\eea
and $p_z = p_0$ and $\sigma_3 = \sigma_0$, one has
\bea
\vec{\sigma}\cdot \hat{p} = 
- \sqrt2 \sigma_-\hat{p}_+ 
+ \sqrt2 \sigma_+ \hat{p}_- + 
\sigma_0\, \hat{p}_0 = \left(
\begin{tabular}{cc}
	$\hat{p}_0$ & $\sqrt2\, \hat{p}_-$ \\
	$-\sqrt2\,\hat{p}_+$ & $-\hat{p}_0$ \\
\end{tabular}
\right)\,,
\eea
where 
$\sigma_\pm^\dagger = \sigma_\mp$ and 
$\hat{p}_\pm^* = - \hat{p}_\mp$.
It leads to 
\bea
\chi^{J=1/2} = 
- \frac{1}{\sqrt3} \vec{\sigma}^T\cdot \hat{p} \chi_C\,,
\eea
with $\chi_C^{T} = (\chi_C^{+},\chi_C^{-})$, 
and 
\bea
\chi_{(J=1/2)}^+ &=& 
\sqrt{\frac23} \hat{p}_+\chi_C^- - \frac{1}{\sqrt3} \hat{p}_- \chi_C^+\,,
\ \ \
\chi_-^{(J=1/2)} = 
-\sqrt{\frac23} \hat{p}_-\chi_C^+ + \frac{1}{\sqrt3} \hat{p}_0 \chi_C^-\,,
\eea
where one can see that the Clebsch-Gordan coefficients of the spin 1/2 state 
from $1\otimes \frac12$ spin states are reproduced. 

\vskip 2mm \noindent
$\bullet$  
$P_{3/2,i}^{(l=1)}$; 
$J^\pi=3/2^+$ state 
for the $\alpha$-$^{13}$C channel
with $l=1$ state and spin-1/2 state

We employ the transition matrices from spin-1/2 to spin-3/2~\cite{fhr-epja12},
\bea
S_1 &=& \frac{1}{\sqrt6}\left(
\begin{tabular}{cccc}
	$-\sqrt3$ & 0 & 1 & 0 \cr
	0 & $-1$ & 0 & $\sqrt3$
\end{tabular}
\right)\,,
\ \ \ 
S_2 = - \frac{i}{\sqrt6}\left(
\begin{tabular}{cccc}
	$\sqrt3$ & 0 & 1 & 0 \cr
	0 & $1$ & 0 & $\sqrt3$
\end{tabular}
\right)\,,
\nnb \\  
S_3 &=& \sqrt{\frac{2}{3}}\left(
\begin{tabular}{cccc}
	$0$ & 1 & 0 & 0 \cr
	0 & $0$ & 1 & $0$
\end{tabular}
\right)\,,
\label{eq;S123}
\eea
with
\bea
S_iS_j^\dagger = \frac23\delta_{ij} - i \frac13 \epsilon_{ijk}\sigma_k\,,
\ \ \ 
\vec{\sigma}\cdot \vec{S} = 0\,. 
\eea
These matrices may be rewritten as 
\bea
S_\pm = \frac12(S_1\pm i S_2)\,, \ \ \ S_0 = S_3
\eea
where 
\bea
S_+ &=& \frac{1}{\sqrt6}\left(
\begin{tabular}{cccc}
	0 & 0 & 1 & 0 \\
	0 & 0 & 0 & $\sqrt3$ 
\end{tabular}
\right)\,,
\ \ \ 
S_- = -\frac{1}{\sqrt6}\left(
\begin{tabular}{cccc}
	0 & 1 & 0 & 0 \\
	$\sqrt3$ & 0 & 0 & 0  
\end{tabular}
\right)\,,
\eea
and one has
\bea
\vec{S}\cdot \hat{l} &=& 
\sqrt2\hat{l}_-S_+ -\sqrt2\hat{l}_+S_+ + \hat{l}_0 S_0
\nnb \\ &=&
\frac{1}{\sqrt3}\left(
\begin{tabular}{cccc}
	$\sqrt3\hat{l}_+$ & $\sqrt2\hat{l}_0$ & $\hat{l}_-$ & 0 \\
	0 & $\hat{l}_+$ & $\sqrt2\hat{l}_0$ & $\sqrt3\hat{l}_-$
\end{tabular}
\right)
\,,
\eea
where $\hat{l}$ is a unit vector, 
and a vector $\vec{l}$ is obtained from the angular momentum
projection operator for $l=1$. 
Now a $J=3/2$ state generated from $l=1$ and $s=1/2$ is obtained as 
\bea
\chi^{(J=3/2)}_m = \vec{S}^T \cdot \hat{l} \chi_C\,,
\eea
where $m=\pm 3/2, \pm 1/2$ and 
\bea
\chi^{(J=3/2)}_{+3/2} &=& \hat{l}_+\chi_C^+\,,
\ \ \ 
\chi^{(J=3/2)}_{+1/2} = 
\frac{1}{\sqrt3}\hat{l}_+\chi_C^-
+ \sqrt{\frac23}\hat{l}_0\chi_C^+ \,,
\nnb \\ 
\chi^{(J=3/2)}_{-1/2} &=& 
\frac{1}{\sqrt3}\hat{l}_-\chi_C^+
+ \sqrt{\frac23}\hat{l}_0\chi_C^- \,,
\ \ \ 
\chi^{(J=3/2)}_{-3/2} = 
\hat{l}_-\chi_C^-\,,
\eea
where one can see that the Clebsch-Gordan coefficients of the spin 3/2 state 
from $1\otimes \frac12$ spin states are reproduced.  

\vskip 2mm 
\noindent
$\bullet$ 
$P_{3/2,ij}^{(l=2)}$; 
$J^\pi=3/2^+$ state 
for the $n$-$^{16}$O system
with $l=2$ and spin-1/2 states 

A naive expectation to construct a projection operator for the $J=3/2$ state
from $l=2$ and spin-1/2 states is to replace the spin-1/2 state to 
the $J=1/2$ state from $l=1$ and spin-1/2 states in the $P_{3/2,i}^{(l=1)}$
operator, $\vec{S}^T\cdot\vec{l}\chi_n 
\to \vec{S}^T\cdot \vec{l}(-1/\sqrt3)\vec{\sigma}^T\cdot\vec{l}\chi_n$. 
To reproduce the Clebsch-Gordan coefficients of the spin 3/2 state from
$2\otimes \frac12$ states, the overall coefficients should be modified.
Thus, we have 
\bea
\chi_m^{(J=3/2)} = - \sqrt{\frac25}\vec{S}^T\cdot\hat{l} 
\vec{\sigma}^T\cdot \hat{l} \chi_n\,.
\eea
Employing the expressions of the Cartesian tensor of rank 2~\cite{sakurai},
\bea
T^{(2)}_{\pm 2} &=& U_{\pm 1}V_{\pm 1} = \hat{l}_\pm \hat{l}_\pm \,,
\ \ \ 
T^{(2)}_{\pm 1} = \frac{1}{\sqrt2}(U_{\pm 1}V_0 + U_0 V_{\pm 1})
= \sqrt2\, \hat{l}_0\hat{l}_\pm \,,
\nnb \\ 
T^{(2)}_0 &=& \frac{1}{\sqrt6} (U_{+1}V_{-1} + 2U_0V_0 + U_{-1}V_{+1}) 
= \sqrt{\frac23}(\hat{l}_+\hat{l}_- + \hat{l}_0\hat{l}_0)\,,
\eea
where vectors $\vec{l}$ are generated from the angular momentum 
operator for $l=2$.
Thus, we have
\bea
\chi^{(J=3/2)}_{+3/2} &=& \sqrt{\frac45} T^{(2)}_{+2}\chi^s_- 
- \frac{1}{\sqrt5} T^{(2)}_{+1}\chi^s_+\,,
\ \ \ 
\chi^{(J=3/2)}_{+1/2} = \sqrt{\frac35} T^{(2)}_{+1}\chi^s_- 
- \sqrt{\frac25} T^{(2)}_{0}\chi^s_+\,,
\nnb \\ 
\chi^{(J=3/2)}_{-1/2} &=& \sqrt{\frac25} T^{(2)}_{0}\chi^s_- 
- \sqrt{\frac35} T^{(2)}_{-1}\chi^s_+\,,
\ \ \ 
\chi^{(J=3/2)}_{-3/2} = \frac{1}{\sqrt5} T^{(2)}_{-1}\chi^s_- 
- \sqrt{\frac45} T^{(2)}_{-2}\chi^s_+\,,
\eea
where one can see that the Clebsch-Gordan coefficients of the spin 3/2 state 
from $2\otimes \frac12$ spin states are reproduced.  

\vskip 2mm \noindent
$\bullet$  
$P_{5/2,ij}^{(l=2)}$; 
$J^\pi=5/2^-$ state 
for the $\alpha$-$^{13}$C channel 
with $l=2$ and spin-1/2 states 

The spin $5/2^-$ state from the $l=2$ and spin-1/2 state 
is constructed as spin-angular functions~\footnote{
	See Eq.~(3.7.64) in Ref.~\cite{sakurai}.
We note that 
$S_i$ matrices in Eq.~(\ref{eq;S123}) can be constructed in the same way. }, 
and we have 
\bea
\frac{1}{\sqrt5} M^T \chi_C\,,
\eea
where $M$ is a $2\times 6$ matrix, 
\bea
M &=& \left(
\begin{tabular}{c c c c c c}
	$\sqrt5 T_{22}$ & $\sqrt4 T_{21}$ & $\sqrt3 T_{20}$ & 
	$\sqrt2 T_{2-1}$ & $T_{2-2}$ & 0 \cr
	0 & $T_{22}$ & $\sqrt2 T_{21}$ & $\sqrt3 T_{20}$ & 
	$\sqrt4 T_{2-1}$ & $\sqrt5 T_{2-2}$
\end{tabular}
\right)\,,
\eea
and $T_{2m}$ ($m=\pm 2,\pm 1,0$) 
are spin-2 states represented as the Cartesian tensor of 
rank 2 as 
\bea
T_{2\pm 2} &=& l_\pm l_\pm\,,
\ \ \
T_{2\pm 1} = \frac{1}{\sqrt2}\left(
l_\pm l_0 + l_0 l_\pm
\right)\,,
\ \ \
T_{20} = \frac{1}{\sqrt{6}}\left(
l_+l_- + 2 l_0l_0 + l_-l_+
\right)\,,
\eea
where we reproduce the Clebsch-Gordan coefficients of the spin 5/2 state 
from $2\otimes \frac12$ spin states. 

Thus, we have
\bea
M &=& \left(
\begin{tabular}{ccc}
	$\sqrt5 l_+l_+$ & $\sqrt2(l_+l_0+l_0l_+)$ & 
	$\frac{1}{\sqrt2}l_+l_- + \sqrt2 l_0l_0 + \frac{1}{\sqrt2}l_-l_+$ 
	\cr
	0 & $l_+l_+$ & $l_+l_0+ l_0l_+$ 
\end{tabular}
\right.
\nnb \\ && \left.
\begin{tabular}{ccc}
	$l_-l_0 + l_0l_-$ & $l_-l_-$ & 0 \cr
	$\frac{1}{\sqrt2}l_+l_- + \sqrt2 l_0l_0 + \frac{1}{\sqrt2} l_-l_+$ &
	$\sqrt2(l_-l_0+ l_0l_-)$ & $\sqrt5 l_-l_-$
\end{tabular}
\right)
\nnb \\ &=&
\left(
\begin{tabular}{cccc}
	$\frac{1}{\sqrt2}l_+$ & $l_0$ & $\frac{1}{\sqrt2}l_-$ & 0 \cr
	0 & $\frac{1}{\sqrt2}l_+$ & $l_0$ & $\frac{1}{\sqrt2}l_-$  
\end{tabular}
\right)
\nnb \\ && \times 
\left(
\begin{tabular}{cccccc}
	$\sqrt{10}l_+$ & $\sqrt4 l_0$ & $l_-$ & & & \cr
	 & $\sqrt2l_+$ & $\sqrt2l_0$ & $l_-$ & & \cr
	 & & $l_+$ & $\sqrt2 l_0$ & $\sqrt2 l_-$ & \cr
	 & & & $l_+$ & $\sqrt4 l_0$ & $\sqrt{10}l_-$ 
\end{tabular}
\right)
\nnb \\ &=& LR \,,
\eea
with
\bea
L = \vec{L}\cdot\vec{l}\,,
\ \ \
R = \vec{R}\cdot \vec{l}\,,
\eea
where
\bea
L_1 &=& -\frac12 \left(
\begin{tabular}{cccc}
	1 & 0 & $-1$ & 0 \cr
	0 & 1 & 0 & $-1$
\end{tabular}
\right)\,,
\\
L_2 &=& -\frac{i}{2} \left(
\begin{tabular}{cccc}
	1 & 0 & $1$ & 0 \cr
	0 & 1 & 0 & $1$
\end{tabular}
\right)\,,
\\
L_3 &=& \left(
\begin{tabular}{cccc}
	0 & 1 & $0$ & 0 \cr
	0 & 0 & 1 & $0$
\end{tabular}
\right)\,,
\\
R_1 &=& \left(
\begin{tabular}{cccccc}
	$-\sqrt5$ & 0 & $1/\sqrt2$ & & & \cr
	 & $-1$ & 0 & $1/\sqrt2$ & & \cr
	 & & $-1/\sqrt2$ & 0 & 1 & \cr
	 & & & $-1/\sqrt2$ & 0 & $\sqrt5$ 
\end{tabular}
\right)\,,
\\
R_2 &=& -i\left(
\begin{tabular}{cccccc}
	$\sqrt5$ & 0 & $1/\sqrt2$ & & & \cr
	 & $1$ & 0 & $1/\sqrt2$ & & \cr
	 & & $1/\sqrt2$ & 0 & 1 & \cr
	 & & & $1/\sqrt2$ & 0 & $\sqrt5$ 
\end{tabular}
\right)\,,
\\
R_3 &=& \left(
\begin{tabular}{cccccc}
	0 & 2 & 0 & & & \cr
	  & 0 & $\sqrt2$ & 0 & & \cr
	  & & 0 & $\sqrt2$ & 0 & \cr
	  & & & 0 & 2 & 0
\end{tabular}
\right)\,.
\eea
Here, the blank spaces in the $R$ matrices represent zero elements.  
We also have relations,
\bea
\vec{\sigma}\cdot \vec{L} = 0\,,
\ \ \
\vec{L}\cdot \vec{R} = 0\,,
\ \ \ 
\vec{\sigma}\cdot L_a \vec{R} = 0\,,
\eea
where $a=1,2,3$. 

\vskip 2mm \noindent
$\bullet$  
$P_{5/2,ijk}^{(l=3)}$; 
$J^\pi=5/2^-$ state 
for the $n$-$^{16}$O channel 
with $l=3$ and spin-1/2 states 

We construct the projection operator of $J=5/2$ from $l=3$ and spin-1/2 
states by imposing a naive expectation. 
We replace the 1/2-spinor acting on 
$P_{5/2,ij}^{(l=2)}$ by $P_{1/2,k}^{(l=1)}\chi_n$, and have
\bea
- \frac{1}{\sqrt{15}} R_i^T L_j^T \sigma_k^T\chi_n \,.
\label{eq;RLsig}
\eea
We note that this expression does not reproduce the terms and coefficients
of the expression of the $J=5/2$ state 
from $3\otimes \frac12$ spin states 
in terms of the Clebsch-Gordan coefficients. 
In the present work, an overall factor of the operator 
can be arbitrary due to the definitions of the neutron width, 
$\Gamma^n_{R(5m)}$.
An arbitrary normalization factor $C$ of the projection operator,
$P^{(l=3)}_{5/2,ijk}$, 
may appear from the vertex function $\Gamma_{dnO}^{(5m)}(p)$, and it is 
absorbed in $\sqrt{\Gamma^n_{R(5m)}}$ in the numerator of the reaction
amplitude.
Another arbitrary factor $C^2$ may appear from the self-energy, 
$\Sigma_{nO}^{(5m)}(p)$, and it is absorbed in $\Gamma^n_{R(5m)}$ in the
denominator. 
The neutron width, $\Gamma^n_{R(5m)}$, should be fitted by experimental data. 
We also confirm that by using the projection operator in Eq.~(\ref{eq;RLsig}), 
the overall factor of the $S$ factor
of the Breit-Wigner formula~\footnote{
	See, e.g., Eq.~(2) in Ref.~\cite{getal-aj12}.
	} 
is reproduced.

\vskip 2mm \noindent
{\bf Appendix B: Angular momentum projection operators 
}

In this appendix, we display the expressions of the projection operators
of the cluster states 
in the relative angular momenta with
$l=0,1,2,3$~\cite{sa-epja16,sa-epja21}. 
For the $\alpha$-$^{13}$C system, 
the angular momentum projection operators, 
$O^{(l=1)}_i$ and $O^{(l=2)}_{ij}$ for $l=1$ and 
$l=2$, respectively,  
are given by  
\bea
O^{(l=1)}_i &=& +i
\left(
 \frac{\stackrel{\leftarrow}{\nabla}}{m_{13C}}
-\frac{\stackrel{\rightarrow}{\nabla}}{m_\alpha} 
\right)_i\,,
\ \ \ 
O^{(l=2)}_{ij} = 
O^{(\alpha C)}_{ij} - \frac13\delta_{ij} O^{(\alpha C)}_{kk}
\,,
\eea
with
\bea
O^{(\alpha C)}_{ij} &=&  
\left(i
 \frac{\stackrel{\leftarrow}{\nabla}}{m_{13C}}
 \right)_i
\left(i
 \frac{\stackrel{\leftarrow}{\nabla}}{m_{13C}}
 \right)_j
-2 
\left(i
 \frac{\stackrel{\leftarrow}{\nabla}}{m_{13C}}
 \right)_i
 \left(i
\frac{\stackrel{\rightarrow}{\nabla}}{m_\alpha} 
\right)_j
+
 \left(i
\frac{\stackrel{\rightarrow}{\nabla}}{m_\alpha} 
\right)_i
 \left(i
\frac{\stackrel{\rightarrow}{\nabla}}{m_\alpha} 
\right)_j
\,.
\label{eq;y1p}
\eea

The angular momentum projection operators of the $n$-$^{16}$O system 
for $l=0,2,3$ are given by 
\bea
O^{(l=0)} &=& 1\,, 
\ \ \ 
O^{(l=2)}_{ij} 
= O^{(nO)}_{ij} - \frac13\delta_{ij}O^{(nO)}_{kk}\,,
\nnb \\
O^{(l=3)}_{ijk} &=&  O^{(nO)}_{ijk} 
- \frac15\left(
\delta_{ij} 
O^{(nO)}_{kll}
+ \delta_{ik} 
O^{(nO)}_{jll}
+ \delta_{jk} 
O^{(nO)}_{ill}
\right)\,,
\eea
with
\bea
O^{(nO)}_{ij} &=& 
\left(i
 \frac{\stackrel{\rightarrow}{\nabla}}{m_O}
 \right)_i
\left(i
 \frac{\stackrel{\rightarrow}{\nabla}}{m_O}
 \right)_j
-2 
\left(i
 \frac{\stackrel{\rightarrow}{\nabla}}{m_O}
 \right)_i
 \left(i
\frac{\stackrel{\leftarrow}{\nabla}}{m_n} 
\right)_j
+
 \left(i
\frac{\stackrel{\leftarrow}{\nabla}}{m_n} 
\right)_i
 \left(i
\frac{\stackrel{\leftarrow}{\nabla}}{m_n} 
\right)_j
\,,
\\
O^{(nO)}_{ijk} &=& 
\left(i
 \frac{\stackrel{\rightarrow}{\nabla}}{m_O}
 \right)_i
\left(i
 \frac{\stackrel{\rightarrow}{\nabla}}{m_O}
 \right)_j
\left(i
 \frac{\stackrel{\rightarrow}{\nabla}}{m_O}
 \right)_k
 -3
\left(i
 \frac{\stackrel{\rightarrow}{\nabla}}{m_O}
 \right)_i
\left(i
 \frac{\stackrel{\rightarrow}{\nabla}}{m_O}
 \right)_j
 \left(i
\frac{\stackrel{\leftarrow}{\nabla}}{m_n} 
\right)_k
\nnb \\ &&
+ 3
\left(i
 \frac{\stackrel{\rightarrow}{\nabla}}{m_O}
 \right)_i
 \left(i
\frac{\stackrel{\leftarrow}{\nabla}}{m_n} 
\right)_j
 \left(i
\frac{\stackrel{\leftarrow}{\nabla}}{m_n} 
\right)_k
-
 \left(i
\frac{\stackrel{\leftarrow}{\nabla}}{m_n} 
\right)_i
 \left(i
\frac{\stackrel{\leftarrow}{\nabla}}{m_n} 
\right)_j
 \left(i
\frac{\stackrel{\leftarrow}{\nabla}}{m_n} 
\right)_k
\,.
\eea

\vskip 2mm \noindent
{\bf Appendix C: Propagators, vertices, and self-energies }

In this appendix, we display and discuss 
the expressions of bare propagators, vertices,
self-energies, and dressed propagators calculated 
from the effective Lagrangian. 
Except for the expressions of the bare propagators, the components
are calculated in the center-of-mass frame. 
The reaction amplitudes of the three resonant states 
of $^{17}$O are calculated by combining those components. 

\vskip 2mm \noindent
{\bf C.1 Bare propagators} 

The inverse of propagators for $n$, $^{16}$O, $\alpha$, $^{13}$C
fields are represented in terms of the standard non-relativistic form,
which are obtained from the effective Lagrangian, 
${\cal L}_0$ in Eq.~(\ref{eq;L0}), as 
\bea
D_P(p_0,\vec{p})^{-1} &=& p_0 - \frac{1}{2m_P}\vec{p}^2 + i\epsilon\,,
\eea
where $P$ represent particles, $P=n,^{16}$O, $\alpha$, $^{13}$C.

The inverse of bare propagators for the resonant $1/2^+$, 
$5/2^-$, $3/2^+$ 
states of $^{17}$O are obtained from 
the effective Lagrangian, ${\cal L}_R$ in Eq.~(\ref{eq;LR}), as 
\bea
\stackrel{\circ}{D}_{S}(p_0,\vec{p})^{-1} &=& \sum_{n=0}^N C_{S}^{(n)} \left(
p_0 - \frac{1}{2m_{S}}\vec{p}^2 + i\epsilon 
\right)^n\,,
\eea
where $S$ represents resonant states, 
$S= 1p, 5m, 3p$. 
In this work, we choose the number of terms 
for the perturbative expansion
as $N=1$ for the $1/2^+$ and $5/2^-$ states and 
$N=3$ for the $3/2^+$ state of $^{17}$O. 
We note that, when one chooses the center-of-mass frame, the total three momenta vanish, $\vec{p}=0$, and the expansion becomes a power series of the energy, 
$p_0^n= E^n$. 

\vskip 2mm \noindent
{\bf C.2 Vertices} 

The $d$-$n$-$^{16}$O vertices for the three resonant states
are obtained from the effective Lagrangian,
${\cal L}_{int}$ in Eq.~(\ref{eq;Lint}), as 
\bea
\Gamma_{dnO}^{(1p)} &=& -y_{1p}'\chi_n^\dagger \,,
\\
\Gamma_{dnO}^{(5m)} &=& 
\frac{1}{\sqrt{15}} \frac{y'_{5m}}{\mu_{nO}^3}
\chi_n^\dagger 
\sigma_a^* L_b^* R_c^* {\cal O}^{(l=3)}_{abc}\,,
\\
\Gamma_{dnO}^{(3p)} &=& 
\sqrt{\frac25} 
\frac{y_{3p}'}{\mu_{nO}^2} 
\chi_n^\dagger
\vec{\sigma}^*\cdot \vec{l}
\vec{S}^*\cdot \vec{l} 
\,,
\eea
with
\bea
{\cal O}_{abc}^{(l=3)} &=& l_a l_b l_c - \frac15\left(
\delta_{ab} l_c + \delta_{ac}l_b + \delta_{bc}l_a
\right)l^2 \,,
\eea
and $l_a$ is the relative momentum of $n$ and $^{16}$O. 

The $d$-$\alpha$-$^{12}$C vertices for the three resonant states 
are obtained from the effective Lagrangian,
${\cal L}_{int}$ in Eq.~(\ref{eq;Lint}), as 
\bea
\Gamma_{d\alpha C}^{(1p)} &=& 
\frac{y_{1p}}{\sqrt3 \mu_{\alpha C}}
\vec{\sigma}^T \cdot \vec{l} \chi_C\,,
\\
\Gamma_{d\alpha C}^{(5m)} &=& - \frac{y_{5m}}{\sqrt5 \mu_{\alpha C}^2} 
\vec{R}^T\cdot \vec{l} \vec{L}^T\cdot \vec{l}\chi_C\,,
\\
\Gamma_{d\alpha C}^{(3p)} &=& - \frac{y_{3p}}{\mu_{\alpha C}}
\vec{S}^T\cdot \vec{l} \chi_C\,,
\eea
where $\vec{l}$ is the relative momentum of $\alpha$ and $^{13}$C.
When we include the initial Coulomb wavefunctions 
for $l=1$ and $l=2$~\cite{sa-epja21}, the vertex functions are obtained as 
\bea
\Psi_{d\alpha C}^{(1p)}(p) &=&  
\frac{y_{1p}}{\sqrt3 \mu_{\alpha C}}
\vec{\sigma}^T \cdot \vec{p} \chi_C \,
e^{i\sigma_1} C_\eta \sqrt{1+\eta^2}
\,,
\\
\Psi_{d\alpha C}^{(5m)}(p) &=& 
-\frac{y_{5m}}{2\sqrt5 \mu_{\alpha C}^2} 
\vec{R}^T\cdot\vec{p} \vec{L}^T\cdot \vec{p} \chi_C
e^{i\sigma_2} \sqrt{(1+\eta^2)(4+\eta^2)} C_\eta\,,
\\
\Psi_{d\alpha C}^{(3p)}(p) &=& 
- \frac{y_{3p}}{\mu_{\alpha C}} 
\vec{S}^T\cdot \vec{p}\chi_C e^{i\sigma_1} 
\sqrt{1+\eta^2}C_\eta\,,
\eea
with
\bea
C_\eta = \sqrt{
	\frac{2\pi\eta}{
		e^{2\pi\eta} - 1
	}
}\,,
\eea
where $p=|\vec{p}|$; 
$\vec{p}$ is the relative momentum of $\alpha$ and $^{13}$C. 
$\eta$ is the Sommerfeld parameter, $\eta = \kappa/p$ where 
$\kappa$ is the inverse of the Bohr radius of the $\alpha$-$^{13}$C system.
$\sigma_1$ and $\sigma_2$ are the Coulomb phase shifts for $l=1$
and $l=2$, respectively. 

\vskip 2mm \noindent
{\bf C.3 Self-energies} 

\vskip 2mm \noindent
$\bullet$ $n$-$^{16}$O loops

The self-energy term from the one-loop diagram of $n$-$^{16}$O 
for the $1/2^+$ state is calculated as 
\bea
i\Sigma^{(1p)}_{nO}(p) &=& 
\left(-iy_{1p}'\right)^2 
\int
\frac{d^4l}{(2\pi)^4} 
\frac{i}{l_0+p_0 - \frac{1}{2m_n}(\vec{l}+\vec{p})^2 + i\epsilon}
\frac{i}{-l_0 - \frac{1}{2m_O}\vec{l}^2 + i\epsilon}
\nnb \\ &=& 2i\mu_{nO}y_{1p}'^2 
\int
\frac{d^3\vec{l}}{(2\pi)^3} 
\frac{1}{
	\vec{l}^2 + D' 
}\,,
\label{eq;Sigma_1p_nO}
\eea
where 
\bea
D' &=& 
	-2\mu_{nO}
	\left(
		E + Q
		\right)
	-i\epsilon
	= - p'^2 -i\epsilon
	\,.
\eea
The loop integral in Eq.~(\ref{eq;Sigma_1p_nO}),  
\bea
I &=& 
\int \frac{d^3\vec{l}}{(2\pi)^3} 
\frac{1}{l^2 + D'}
= \frac{1}{2\pi^2}\left(
\int_0^\infty dl 
- \frac{\pi}{2}\sqrt{D'}
\right)
\,,
\eea
has a linear divergence, and  
it is renormalized by the term, $C_{1p}^{(0)}$. 
The finite part of the self-energy of the $n$-$^{16}$O loop for 
the $1/2^+$ state is obtained as 
\bea
\Sigma_{nO}^{(1p)}(p) &=& 
-y_{1p}'^2 \frac{\mu_{nO}}{2\pi}
\sqrt{D'}
= +y_{1p}'^2 \frac{\mu_{nO}}{2\pi} (i p')
\,.
\eea

The self-energy term from the $n$-$^{16}$O loop for the $5/2^-$ 
state is 
\bea
i\Sigma_{nO}^{(5m)}(p) &=& i^4 
\frac{1}{15} \frac{y_{5m}'^2}{\mu_{nO}^6} 
\int \frac{d^4l}{(2\pi)^4}
\frac{
	R^T_aL^T_b\sigma^T_c \sigma^*_i L^*_j R^*_k 
	\hat{O}_{abc}^{(l=3)}
	\hat{O}_{ijk}^{(l=3)}
	}{
		(l_0 + E + Q -\frac{1}{2m_n}\vec{l}^2 + i\epsilon )
		(-l_0 - \frac{1}{2m_O}\vec{l}^2 + i\epsilon)
	}
	\nnb \\ &=&
	i
\frac{y_{5m}'^2}{15\mu_{nO}^6}
\frac{2\mu_{nO}}{2\pi^2} 
\int_0^\infty dl \frac{l^8}{l^2+D'} 
\int\frac{d\Omega_{\hat{l}}}{4\pi} 
	R^T_aL^T_b\sigma^T_c \sigma^*_i L^*_j R^*_k 
	\hat{O}_{abc}^{(l=3)}
	\hat{O}_{ijk}^{(l=3)}
	\,,
\eea
where
\bea
\hat{O}_{abc}^{(l=3)} &=& \hat{l}_a\hat{l}_b\hat{l}_c 
- \frac15\left(\delta_{ab}\hat{l}_c + \delta_{ac}\hat{l}_b 
+ \delta_{bc}\hat{l}_a
\right)\,,
\eea
and 
\bea
\int_0^\infty dl \frac{l^8}{l^2+D'} &=& 
\int_0^\infty dl\left(
l^6 - D'l^4 + D'^2 l^2 - D'^3 + \frac{D'^4}{l^2+D'}
\right)\,,
\eea
and the four divergent terms are renormalized by the terms,
$C_{5m}^{(3)}$, $C_{5m}^{(2)}$, $C_{5m}^{(1)}$, $C_{5m}^{(0)}$,
and the finite part of the integral is obtained as 
\bea
\int_0^\infty dl \frac{D'^4}{l^2+D'} = \frac{\pi}{2} D'^{7/2}\,.
\eea
The angle integral is calculated as 
\bea
\int\frac{d\Omega_{\hat{l}}}{4\pi} 
R_a^\dagger L_b^\dagger \sigma_c^\dagger \sigma_d L_e R_f 
\hat{O}_{abc}^{(l=3)} \hat{O}_{def}^{(l=3)} &=& \frac23I_{6\times 6}\,,
\eea
where $I_{6\times 6}$ is the $6\times 6$ unit matrix, 
and we have the finite part of the self-energy of the $n$-$^{16}$O loop
for the $5/2^-$ state as 
\bea
\Sigma_{nO}^{(5m)}(p) &=& 
\frac{2y_{5m}'^2}{45\mu_{nO}^6} \frac{\mu_{nO}}{2\pi} 
D'^{7/2}
=\frac{2y_{5m}'^2}{45\mu_{nO}^6} \frac{\mu_{nO}}{2\pi} (ip'^7)
\,.
\eea

The self-energy from the $n$-$^{16}$O loop 
for the $3/2^+$ state is
\bea
i\Sigma_{nO}^{(3p)}(p) &=& 
+i\frac25 \frac{\mu_{3p}'^2}{\mu_{nO}^4}(2\mu_{nO})
\int \frac{d^3\vec{l}}{(2\pi)^3} 
\frac{
	\vec{S}^T\cdot\vec{l}\vec{\sigma}^T\cdot \vec{l}
	\vec{\sigma}^*\cdot \vec{l}
	\vec{S}^*\cdot\vec{l}
	}{
		\vec{l}^2 + D'
	}\,,
	\label{eq;Sigma_3p_nO}
\eea
and the integral in the self-energy term is calculated as 
\bea
I &=& 
\int \frac{d^3\vec{l}}{(2\pi)^3} 
\frac{
	\vec{S}^T\cdot\vec{l}\vec{\sigma}^T\cdot \vec{l}
	\vec{\sigma}^*\cdot \vec{l}
	\vec{S}^*\cdot\vec{l}
	}{
		\vec{l}^2 + D'
	}
	\nnb \\ &=& 
	\frac{1}{2\pi^2} 
	\int_0^\infty dl \frac{l^6}{l^2+D'} 
	\int \frac{d\Omega_{\hat{l}}}{4\pi}
	\vec{S}^T\cdot\hat{l}\vec{\sigma}^T\cdot \hat{l}
	\vec{\sigma}^*\cdot \hat{l}
	\vec{S}^*\cdot\hat{l}
	\,,
\eea
where the part of the momentum integral leads to 
\bea
I_l &=& \int_0^\infty dl \frac{l^6}{l^2+D'} 
=
\int_0^\infty dl\left(
l^4 - D'l^2 + D'^2 - \frac{D'^3}{l^2+D'}
\right)\,.
\eea
The three divergent terms are renormalized by 
the terms, $C_{3m}^{(2)}$, $C_{3m}^{(1)}$, $C_{3m}^{(0)}$, 
and the finite part is obtained as 
\bea
I_l^{fin} &=& - \frac{\pi}{2} D'^{5/2}\,.
\eea
By using the relation of the angle integral
\bea
\int\frac{d\Omega_{\hat{l}}}{4\pi} 
\hat{l}_a\hat{l}_b\hat{l}_c \hat{l}_d
&=& 
\frac{1}{15}\left(
\delta_{ab}\delta_{cd} + \delta_{ac}\delta_{bd} + \delta_{ad}\delta_{bc}
\right)
\,,
\eea
one has
\bea
I_{\hat{l}} &=& 
	\int \frac{d\Omega_{\hat{l}}}{4\pi}
	\vec{S}^T\cdot\hat{l}\vec{\sigma}^T\cdot \hat{l}
	\vec{\sigma}^*\cdot \hat{l}
	\vec{S}^*\cdot\hat{l}
	=
	\frac{1}{15}\left(
	\vec{S}^\dagger \cdot \vec{\sigma}^\dagger \vec{\sigma}\cdot \vec{S}
	+ S_i^\dagger \sigma_j^\dagger \sigma_i S_j
	+ S_i^\dagger \sigma_j^\dagger \sigma_j S_i
	\right)^*
	\nnb \\ &=& 
	\frac{1}{15} \left[
	S_i^\dagger (\delta_{ji} + i \epsilon_{jik}\sigma_k) S_j
	+ 3 \vec{S}^\dagger\cdot \vec{S}
	\right]
= 
\frac13 \vec{S}^\dagger \cdot \vec{S}
= \frac13 I_{4\times 4}\,,
\eea
where 
$i\epsilon_{jik}S_i^\dagger \sigma_kS_j = I_{4\times 4}$, and 
$I_{4\times 4}$ is the 4$\times$4 unit matrix. 
Thus, we have
\bea
\Sigma_{nO}^{(3p)}(p) &=& 
- \frac{2}{15} \frac{y_{3p}'^2}{\mu_{nO}^4} 
\frac{\mu_{nO}}{2\pi} 
D'^{5/2}
= \frac{2}{15} \frac{y_{3p}'^2}{\mu_{nO}^4} 
\frac{\mu_{nO}}{2\pi} (ip'^5)
\,.
\eea

\vskip 2mm \noindent
$\bullet$ $\alpha$-$^{13}$C loops

The self-energy term from the $\alpha$-$^{13}$C loop for the $1/2^+$
state is obtained as~\cite{sa-epja21}
\bea
\Sigma_{\alpha C}^{(1p)}(p) &=& 
\frac{y_{1p}^2}{3\mu_{\alpha C}^2}
\left(
\frac{\mu_{\alpha C}}{2\pi} 
2\kappa H_1(p)
\right)\,,
\eea
where the function $H_1(p)$ is given in Eq.~(\ref{eq;H1}).

The self-energy term of $\alpha$-$^{13}$C loop for the $5/2^-$ state
is obtained as 
\bea
\Sigma_{\alpha C}^{(5m)}(p) &=& 
\frac{2y_{5m}^2}{15\mu_{\alpha C}^4} \left(
\frac{\mu_{\alpha C}}{2\pi} 2\kappa H_2(p) 
\right)\,,
\eea
where we used the relations,
\bea
R^\dagger_a L^\dagger_b L_xR_y
\frac{1}{15}\left(
\delta_{ax}\delta_{by} 
+ \delta_{ay}\delta_{bx}
- \frac23\delta_{ab}\delta_{xy} 
\right) = \frac23I_{6\times 6}\,,
\eea
with
\bea
R^\dagger_a L^\dagger_b L_b R_a = 
R^\dagger_a L^\dagger_b L_a R_b = 5 I_{6\times 6}\,, 
\ \ \ 
\vec{L}\cdot \vec{R} = 0\,,
\eea
and $I_{6\times 6}$ is the 6$\times$6 unit matrix. 

The self-energy term of $\alpha$-$^{13}$C loop 
for the $3/2^+$ resonant state of $^{17}$O 
is obtained as 
\bea
\Sigma_{\alpha C}^{(3p)}(p) &=&
\frac{y_{3p}^2}{3\mu_{\alpha C}^2} 
\left(
\frac{\mu_{\alpha C}}{2\pi}
2\kappa H_1(p)
\right)\,.
\eea

\vskip 2mm \noindent
{\bf C.4 Dressed propagators} 

The inverse of dressed propagators of the 
resonant states are obtained, 
as presented Feynman diagrams in Fig.~\ref{fig;propagators},
by including the self-energy terms. 
We obtain them in the center-of-mass frame as 
\bea
D_{S}(p)^{-1} = \sum_{n=0}^N C_S^{(n)} E^n 
+ \Sigma_{\alpha C}^{(S)}(p) 
+ \Sigma_{nO}^{(S)}(p) \,,
\eea
where $S$ are labels for the resonant $1/2^+$, $5/2^-$, $3/2^+$
states of $^{17}$O with $S = 1p,5m,3p$, 
and $p$ is the magnitude of the relative momentum, $p=|\vec{p}|$,
and $E=p_0= p^2/(2\mu)$. 

Thus, the inverse of the dressed propagators of the resonant states
of $^{17}$O in the center-of-mass frame are composed as 
\bea
D_{1p}(p)^{-1} &=& 
\sum_{n=0}^N C_{1p}^{(n)} p_0^n 
+ \Sigma_{\alpha C}^{(1p)}(p) 
+ \Sigma_{nO}^{(1p)}(p)
\nnb \\ &=& 
\sum_{n=0}^N C_{1p}^{(n)} E^n
+ \frac{y_{1p}^2}{3\mu^2} 
\left(
\frac{\mu}{2\pi} 
2\kappa H_1(p)
\right)
+ y_{1p}'^2 \frac{\mu'}{2\pi} (ip')
\,,
\\
D_{5m}(p)^{-1} &=& 
\sum_{n=0}^N C_{5m}^{(n)} E^n 
+ \Sigma_{\alpha C}^{(5m)}(p) 
+ \Sigma_{nO}^{(5m)}(p) 
\nnb \\ 
&=& 
\sum_{n=0}^N C_{5m}^{(n)} E^n 
+ \frac{2y_{5m}^2}{15\mu^4}
\left(
\frac{\mu}{2\pi}
2\kappa H_2(p) 
\right)
+ \frac{2y_{5m}'^2}{45\mu'^6} 
\frac{\mu'}{2\pi} 
(ip'^7)
\,,
\\
D_{3p}(p)^{-1} &=& \sum_{n=0}^N C_{3p}^{(n)} p_0^n 
+ \Sigma_{\alpha C}^{(3p)}(p) 
+ \Sigma_{nO}^{(3p)}(p) 
\nnb \\ &=& 
\sum_{n=0}^N C_{3p}^{(n)} E^n
+ \frac{y_{3p}^2}{3\mu^2} 
\left(\frac{\mu}{2\pi} 2\kappa H_1(p)\right)
+ \frac{2}{15} 
\frac{y_{3p}'^2}{\mu'^4} 
\frac{\mu'}{2\pi}
(ip'^5)
\,,
\eea
where $E$ is the energy of the $\alpha$-$^{13}$C system in the center-of-mass
frame, 
$p=\sqrt{2\mu E}$, 
$p' = \sqrt{2\mu'(E+Q)}$,
and $Q$ is the $Q$-value,
$Q=2.22$~MeV.  

\end{document}